\documentclass[fleqn,useAMS,usenatbib]{mnras}

\usepackage[dvips]{graphics}
\usepackage{graphicx}
\usepackage[latin1]{inputenc}

\usepackage[T1]{fontenc}
\usepackage{ae,aecompl,url}
\usepackage{multicol}

\newcommand{\appropto}{\mathrel{\vcenter{
  \offinterlineskip\halign{\hfil$##$\cr
    \propto\cr\noalign{\kern2pt}\sim\cr\noalign{\kern-2pt}}}}}

\title[]{Obliquity evolution of the minor satellites of Pluto and Charon}

\author[Quillen et al.]
{
Alice C. Quillen$^{1}$, 
Fiona Nichols-Fleming${^1}$,  
Yuan-Yuan Chen$^{1,2}$
and Beno\^it Noyelles$^3$
\\
$^1$Department of Physics and Astronomy, University of Rochester, Rochester, NY 14627 USA \\
$^2$Key Laboratory of Planetary Sciences, Purple Mountain Observatory, Chinese Academy of Sciences, Nanjing 210008, China\\
$^3$Department of Mathematics and the Namur Centre for Complex Systems (naXys), \\
\ University of Namur, 
8 Rempart de la Vierge, Namur B-5000 Belgium
}

\begin{document}
\maketitle

\begin{abstract}

New Horizons mission observations show that
the small satellites Styx, Nix, Kerberos and Hydra,
of the Pluto-Charon system, have not tidally spun-down to near
synchronous spin states and have high obliquities with respect to their orbit  
about the Pluto-Charon binary \citep{weaver16}.
We use a damped mass-spring model within an N-body simulation 
to study  spin and obliquity evolution for single spinning 
non-round bodies in circumbinary orbit.
Simulations with tidal dissipation alone do not show strong obliquity variations from tidally induced
spin-orbit resonance crossing 
and this we attribute to the high satellite spin rates and low orbital eccentricities.
However, a tidally evolving
Styx exhibits intermittent obliquity variations and episodes of tumbling.
During a previous epoch where Charon migrated away from Pluto,
the minor satellites could have been trapped in 
orbital mean motion inclination resonances.  An outward migrating Charon induces
large variations in Nix and Styx's obliquities.
The cause is a commensurability between the mean motion resonance frequency
and the spin precession rate of the spinning body.
As the minor satellites are near mean motion resonances,
this mechanism could have lifted the obliquities of all four minor satellites.
The high obliquities need not be primordial if 
the minor satellites were at one time captured into mean motion resonances.
  
\end{abstract}

\vskip 0.05 truein
\vskip 0.05 truein

\section{Introduction}

The  five satellites of Pluto are Charon, Styx, Nix, Kerberos, and Hydra, in order of distance from Pluto
\citep{weaver06,showalter11,showalter12}. 
The satellite system is nearly coplanar with orbital periods near ratios of 1:3:4:5:6, 
but sufficiently distinct from integer ratios relative to Charon's orbital period
 that the small satellites are not presently in mean-motion resonances with Charon (e.g., \citealt{buie13}).   
As the masses of Pluto and Charon vastly exceed the
masses of the other satellites, we refer to Pluto and Charon as a binary (following \citealt{stern92})
and Styx, Nix,  Kerberos and Hydra as minor satellites of the Pluto-Charon binary.

Over 1--10 Myr, tidal evolution should have synchronized the rotation of Pluto and Charon 
 and then circularized their orbit,
 expanding the binary to its present separation \citep{farinella79}. 
 Tidal evolution of Pluto-Charon would lead to capture of the minor satellites
 into mean motion orbital resonances.  
 However numerical integrations have shown that this often 
 causes such wide-scale dynamical instability that  
resonant transport (migration) of 
minor satellites to their current location probably did not take place  \citep{cheung14}.
Alternatively the smaller satellites could have formed from a circumbinary disk, and after
 the formation  
 of the   Pluto-Charon binary \citep{lithwick08,kenyon14}.
 
Prior to the arrival of the New Horizons mission at Pluto, \citet{showalter15}  explored possible
 spin states for the minor satellites.   They speculated that
the minor satellites would have tidally spun down and so would be 
 slowly spinning, with angular spin rate $w$ a similar size as
  the orbital mean motion $n_o$. 
 \citet{showalter15}  speculated that the minor satellites would be chaotically wobbling or tumbling 
due to instability associated with spin-orbit resonances 
\citep{colombo65,goldreich66,wisdom84,celletti90,melnikov10} 
(resonances where $2w$ is a multiple of $n_o$). 
 \citet{showalter15} also suggested  that
perturbations from Charon could affect the rotation state
of the minor satellites, contributing to chaotic tumbling, 
in analogy to how an orbital resonance between Titan and Hyperion could affect
the rotational state of Hyperion.  
A dynamical mechanism for the chaotic behavior was proposed
 by \citet{correia15} who showed that for slowly spinning satellites,
spin-binary resonances from Charon's periodic perturbations are sufficiently strong to
cause chaotic tumbling.  
However, New Horizons observations showed that the low mass satellites are spinning faster
 than considered by \citet{showalter15,correia15},
with angular spin rates $w \ga 6 n_o$,  many times greater than 
their orbital mean motions, implying that despinning due to tidal dissipation
has not taken place \citep{weaver16}.  
At higher spin states, spin-orbit and spin-binary resonances may not be
as strong, so chaotically tumbling is not assured.

\subsection{Obliquities}

The obliquity is the angle between the spin vector of satellite and its orbit normal.
For the four minor satellites,
\citet{weaver16} measured the angle between minor satellite spin vector and
the  orbit normal of the Pluto-Charon binary and at low orbital inclination this angle
is equal to the obliquity.
\citet{weaver16} 
found that all four minor satellites have obliquities near 90 degrees, with spin vector 
lying nearly in their orbital planes.  

Are the high obliquities surprising?  If the minor satellites have not tidally spun-down then
tidal evolution may not have varied their obliquities.   
 The high obliquities could be primordial and acquired during
formation.   The minor satellite orbital inclinations and eccentricities are all low, with all
inclinations below 0.5$^\circ$ and
eccentricities all below that of Hydra at 0.00554 \citep{brozovic15}.
The low orbital inclinations and eccentricities suggest that the minor satellites were
formed in situ, in a circumbinary disk.
Postulated is minor satellite formation  in ring of debris comprised of 
material ejected during 
the impact that formed the Pluto-Charon binary (see \citealt{kenyon14}).

Perhaps we can compare the obliquities of  Pluto and Charon's minor
satellites with other satellites that have high spin rates.
Among satellites with known rotation states
\cite{melnikov10} found only seven rapidly rotating satellites, with
periods less than a day.
All of them are irregular satellites, 
with a possible exception of Nereid \citep{sheppard06}.
\citet{peale77} showed that irregular satellites
should reside close to their initial (and rapid) rotation states.
We make the distinction between regular and irregular satellites as
an irregular satellite can be a captured object whereas a regular satellite 
could be formed in a circum-planetary disk.   We would expect
the primordial spin of a captured object to be randomly oriented
whereas a regular satellite, like a planet, could be formed at zero obliquity
(we will discuss this assumption for Pluto and Charon's minor satellites below).
Nereid with eccentricity 0.75 has a low inclination ($7^\circ$ with respect
to Neptune's Laplace plane).  However, a multiple year photometric study of Nereid's
light curve \citep{shaefer08} suggests that the rotation rate is not as
fast as originally measured (at 0.48 days; \citealt{grav03}) but could be
much slower and the body could be tumbling.  Even if Nereid was originally a regular
satellite, it may not currently be rapidly spinning.

The majority of regular planetary satellites with known rotation states rotate synchronously.
The remaining regular satellites with known rotation states 
are tumbling with spin similar in magnitude (within a factor of a few)
as the orbital mean motion \citep{peale77,melnikov10}.  
Thus the solar system lacks rapidly spinning {\it regular} satellites.
Placed in this context the minor satellites of
Pluto and Charon are unique (though see \citealt{hastings16} on Haumea's satellites).  
Styx, Nix, Kerberos and Hydra are similar to regular satellites of planets
as they have low eccentricities and inclinations, yet their spin periods
are much shorter than their orbital periods and so they can be considered
rapid rotators (as classified by \citealt{melnikov10}).     

What primordial obliquity distribution is predicted for Pluto and Charon's minor satellites?
A proto-planet  that forms in a disk via accretion of gas and small planetesimals is expected
to form at low obliquity \citep{lissauer91,johansen10}.  However, if planets 
accrete at late stages
from a distribution of massive non-interacting planetesimals, their obliquity distribution
can be consistent random spin orientations \citep{dones93,kokubo07}.
The low inclinations and eccentricities of Pluto and Charon's minor satellites
and inferred epoch of radial migration 
suggest that a circumbinary ring of small particles was present after minor 
satellite formation \citep{kenyon14} and implying that some fraction of accreted
material incorporated into these satellites originated from low mass particles.
We cannot rule out the possibility that the primordial spin states were randomly oriented, 
but neither is this assured.

We consider the possibility that four randomly oriented spinning
bodies have obliquity distribution similar to Pluto and Charon's minor satellites.
\citet{weaver16} measured obliquities of $\epsilon = $ 91, 123, 96, 110$^\circ$ for Styx, Nix, Kerberos
and Hydra, respectively, with an estimated error of $\pm 10^\circ$.
The probability distribution for randomly oriented spin directions 
 peaks  at an obliquity of $90^\circ$ and obeys
 probability distribution $P(\epsilon) = \frac{1}{2} \sin \epsilon$.  
 All four obliquities are within an equatorial
 band $\epsilon \in [90,123]$.  Integrating the probability distribution between
 the two boundaries gives a probability of $0.5\cos 123^\circ = 0.272$.
The probability that 4 randomly oriented objects are all found in this same band 
is $P \sim 0.272^4 = 0.0055$.   In an ensemble of 1000  systems of four randomly oriented minor
satellites  only 6 systems would
have all four satellite obliquities  in the band between 90 and 123$^\circ$.
Pluto and Charon's minor satellites are a single system so this is not a meaningful
 statistical result. Nonetheless the low probability does motivate study of mechanisms for
tilting the spin axes after formation, away from their primordial values.

\subsection{Evolution of spin states}
 
 As a satellite despins due to tidal dissipation, 
 it may be captured in spin-orbit resonant states
  \citep{colombo65,goldreich66,peale77,wisdom84,celletti}.  
 However a body that is only very slowly spinning down due to tidal dissipation 
 could cross  spin-orbit resonances or spin-binary resonances if there is a drift in
 the satellite's semi-major axis, known as `orbital migration'.   
  Attitude instability, leading to obliquity variations and chaotic behavior, is common within
spin-orbit resonance  \citep{wisdom84,melnikov08,melnikov10} and expected
in spin-binary resonance \citep{correia15}.
There may be a connection between the minor satellite obliquities and previous
 episodes of spin-orbit or spin-binary resonance crossing or capture.
 
External to spin-orbit resonance, 
tidal dissipation causes a spherical body initially at low obliquity 
 and $w/n_o \ga 6$  to slowly increase in obliquity \citep{goldreich70,ward75,gladman96},
 unless the viscoelastic relaxation timescale and eccentricity are both high \citep{boue16}. 
However, the obliquity drift rate due to tidal dissipation is slower than but a similar size as
 the tidal spin-down rate \citep{goldreich70}.  
Our numerical integrations 
have confirmed that this remains true for elongated non-spherical bodies.  
If the minor satellites have not spun down,
 then neither should their obliquities have approached $90^\circ$. 
As long as minor satellite primordial obliquities were not all near $90^\circ$, then
it is unlikely that the current near $90^\circ$
minor satellite obliquities in the Pluto-Charon system are due to tidal secular  (non-resonant)
obliquity evolution alone.  

With near integer orbital period ratios between satellites,
the Pluto-Charon satellite system is 
near orbital mean motion resonances and may have crossed or been captured
into these resonances in the past.  Migration could have take place due to tidal evolution
of Pluto and Charon but also due to interactions with a previous and now absent
circumbinary disk \citep{lithwick08,cheung14,kenyon14}. 
The minor satellites themselves could have been embedded in a disk and migrated by
 driving spiral density waves into the disk.  Inwards or outwards migration could have taken place
 (e.g., \citealt{lubow}).    

 Planetary orbits have small inclinations and undergo precession (a rate of change
 of the longitude of the ascending node) due to mutual planet/planet gravitational perturbations.
 A similarity in a body's spin precession rate and a precession rate 
  of its orbit or the orbit of a perturber can cause obliquity variations,
 \citep{ward04,correia16}, a mechanism described as `secular spin-orbit resonance'.
Secular spin-orbit resonant mechanisms may have operated 
 in the Pluto-Charon system.    Capture in a mean motion resonance with the Pluto-Charon
 binary  can lift the orbital inclination of a minor satellite and perturb its orbital precession
 frequencies.    The obliquities of the minor satellites could
have been influenced by past proximity or capture in the mean motion resonances.

To investigate these mechanisms and their possible role in influencing the obliquities of
Pluto and Charon's minor satellites we carry out a numerical study of the spin and obliquity
evolution of a non-spherical body in orbit about a binary system.
 Because of their simplicity and speed, compared to more computationally intensive grid-based or finite element methods, mass-spring computations are an attractive method for simulating deformable bodies.
By including spring damping forces they can model viscoelastic tidal deformation.
We previously used a mass-spring model to study tidal encounters  \citep{quillen16a}, 
measure tidal spin down for spherical bodies over a range
of viscoelastic relaxation timescales  \citep{frouard16}, and spin-down of triaxial bodies 
spinning about a principal body axis aligned with the orbit normal \citep{quillen16b}.
 Here we use the same type of simulations to study longer timescale obliquity and
 spin evolution.
Like \citet{mardling02,boue09}, we compute torques on  spinning bodies that are in orbit about
point masses. Rather than averaging
over body shape or orbit, we can take into account viscous dissipation in the extended simulated
body directly 
(using our damped springs).  The four minor satellites are not
round so we simulate bodies with body axis ratios
based on observed and measured values. 
For other simulation techniques integrating orbits and body rotation see 
\citet{showalter15,correia15,correia16,hou16,boue17}. 

Before we begin our numerical study we tabulate and estimate parameters for 
Pluto and Charon and their minor satellites.  
We reexamine estimates for the tidally induced spin-down time, then estimate
 the wobble decay time
and an asphericity parameter used to characterize the strength or libration
frequency of spin-orbit resonances.
We also compute precession frequencies for the spin axis and the orbital longitude of
the ascending node.  These parameters will help us interpret our simulations.
Our simulations are described in section \ref{sec:sims}. Simulations with tidal dissipation alone
are described in section \ref{sec:nd} and those allowing the binary to drift in section \ref{sec:bin}.
A possible mechanism for lifting the minor satellite obliquities (from initially low values) 
is identified from the simulations
and discussed further in our final section \ref{sec:sum}.

\begin{figure}
\includegraphics[width=3.1in]{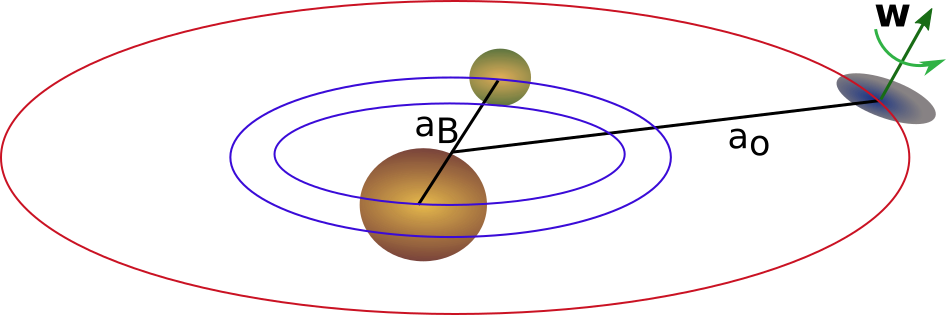}
\caption{We simulate a resolved spinning satellite about the Pluto-Charon binary.
Pluto and Charon are modeled as point masses whereas the minor satellite is modeled
with masses and springs.
\label{fig:bin}}
\end{figure}

\section{Parameters, time-scales and frequencies}

\subsection{Tidal spin-down times}


Spin-down timescales are often computed for spherical bodies in orbit about a single mass.
We start with a spherical body of radius $R$, mass $M$ in orbit with orbital semi-major axis
$a_o$ about a body of mass $M_*$.   We use $a_o$ to denote orbital
semi-major axis and $a$ to denote body semi-major axis.
After discussing the tidal spin-down timescales for a
spherical body in orbit about a point mass,  we will consider
non-spherical bodies described by the radius of the equivalent volume sphere, $R_v$,
and in orbit about a binary rather than single mass.  

The secular part of the semi-diurnal ($l=2$) term in the Fourier expansion of the perturbing
potential (e.g., see the appendix by \citealt{frouard16})
gives a tidally induced torque on the spherical body
\begin{equation}
T = \frac{3}{2a_o} G M_*^2 \frac{R^5}{a_o^5}  k_2 (\sigma_t) \sin \epsilon_2(\sigma_t) 
\end{equation}
(also see \citealt{kaula64,goldreich63,goldreich68,M+D,efroimsky13}),
with $G$ the gravitational constant.
Here the tidal frequency $\sigma_t = 2 (n_o - \dot \vartheta) = 2 (n_o - w)$
where $n_o$ is the orbital mean motion.
When spinning about a principal axis, with spin axis perpendicular to the orbital
plane and $\vartheta$ giving the orientation angle of principal body axis 
in the orbital plane,   the angular spin of the body $\dot \vartheta = w$.
The quality function is $k_2  (\sigma_t) \sin (\epsilon_2 (\sigma_t))$
and is often approximated as $2 k_2/Q$ with $Q$ a tidal dissipation factor   
(e.g., \citealt{kaula64}) and $k_2$ a Love number.

The Love number for an incompressible homogeneous elastic body
\begin{equation}
k_2 = \frac{3/2}{1 + \tilde \mu} \label{eqn:k2}
\end{equation}
with 
\begin{equation}
\tilde \mu = \frac{19 \mu*}{2 \rho g R}  = \frac{38\pi}{3} \frac{\mu*}{e_g} \label{eqn:tildemu}
\end{equation}
\citep{M+D,burns77}
where $\rho$ is the mean density, 
$g =  GM/R^2$ is the gravitational acceleration at the surface, and
$\mu*$ is the elastic shear modulus or rigidity.   We use 
\begin{equation}
e_g \equiv \frac{GM^2}{R^4} = 1.2 {\rm GPa} \left(\frac{R}{1000 {\rm km}} \right)^2
 \left(\frac{\rho}{1 {\rm g~cm}^{-3}} \right)^2
  \label{eqn:eg} 
\end{equation}
for a unit of central pressure or gravitational energy density.
It is common to estimate $\mu* \approx 4 \times 10^9 {\rm N\ m}^{-2} = 4~{\rm GPa}$ 
for icy bodies (e.g., see \citealt{nimmo06,M+D}).
Inserting equation \ref{eqn:tildemu} into equation \ref{eqn:k2} for a small icy body 
($R < 1000$ km; $\tilde \mu >1$)
gives
\begin{equation}
k_2 \approx 0.038 \frac{e_g}{\mu*}.  \label{eqn:k2b}
\end{equation}
The spin-down time can be estimated from the body's moment of inertia, $I$, and
an initial spin $w_{init}$ giving $t_{despin} \sim I w_{init} /T$ or
\begin{equation}
t_{despin} \approx \frac{I w_{init}a_o}{3 GM_*^2} \left( \frac{a_o}{R}\right)^5 \frac{Q}{k_2}
\label{eqn:tdespin}
\end{equation}
(see equation 9 by \citealt{gladman96}; \citealt{peale77}).
Using the moment of inertia for a spherical body, $I = \frac{2}{5} M R^2$,
and setting the initial spin to be that of centrifugal breakup 
$w_{init} = \sqrt{GM/R^3}$, the spin-down time
\begin{equation}
\frac{t_{despin}}{P_o} \approx  \frac{1}{15 \pi} \left(\frac{M}{M_*} \right)^\frac{3}{2}
\left( \frac{a_o}{R} \right)^\frac{9}{2} \frac{Q}{k_2}, \label{eqn:tlock}
\end{equation}
where $P_o$ is the orbital period.
Here the spin-down time is estimated for a spherical body with prograde
spin axis perpendicular to the orbit plane.

To approximate the spin down times for non-spherical bodies,
we replace the body radius with the radius of the equivalent volume sphere, $R_v$
(see \citealt{quillen16b}). 
It is convenient to define a gravitational timescale
 \begin{eqnarray}
t_g &=& \sqrt{ \frac{ R_v^3}{GM}}  = \sqrt{\frac{3}{4\pi G \rho}}   \label{eqn:tg} \\
&=& 2000\ {\rm s} \left( \frac{\rho}{ 1 \ {\rm g \ cm}^{-3}} \right)^{-\frac{1}{2}}. \nonumber 
\end{eqnarray}
Following \cite{quillen16b}
our parameter $e_g$ (equation \ref{eqn:eg}) for non-spherical bodies  has
radius $R$ replaced by $R_v$.

To estimate spin down times for objects in orbit about a binary rather than single
mass, we replace $M$ with $M_B$, the total mass of the binary.
The orbital semi-major axis $a_o$ of the
spinning body is computed with respect to the center of mass of the binary (see Figure \ref{fig:bin}) and using the total binary mass.  It can be called an {\it osculating} orbital element \citep{renner06}.  The osculating mean motion $n_o \equiv \sqrt{GM_B/a_o^3}$.

We  tabulate properties of the Pluto-Charon binary in Table \ref{tab:PC}.  
Values for the binary or Charon's period,  $P_C$, binary or Charon's semi-major axis, $a_C$, 
mass of the binary, Charon to Pluto mass ratio, and binary reduced mass are
based on those measured by \citet{brozovic15}. 

\begin{table}
\vbox to80mm{\vfil
\caption{\large  Pluto-Charon binary  \label{tab:PC}}
\begin{tabular}{@{}lllllll}
\hline
$G M_{PC}$ & $975.5 \pm 1.5$ km$^3$ s$^{-2}$ \\
$G M_C$ & $105.88 \pm 1.0$ km$^3$ s$^{-2}$ \\
$M_C/M_P$ & 0.12 \\
$\mu_{PC}/M_{PC}$ & 0.0967  \\
$P_C$  & 6.3872 days \\ 
$a_C$  & 19596 km \\
$\mu_B a_C^2 /m_{PC}$ &  $3.715 \times 10^7$ km$^2$\\
\hline 
\end{tabular}
{\\ Here $\mu_{PC}$ is the reduced mass of Pluto-Charon binary and $M_{PC}$ is 
the sum of Pluto and Charon's masses.      $M_C$ is the mass of Charon.
We computed the ratio $\mu_B/M_{PC}$ using values listed in Table 10 by \citet{brozovic15}.  
Here $P_C$ is the rotation period of the Pluto-Charon binary
 from Table 9 by \citet{brozovic15}.
 $GM_{PC}$ is the gravitational constant times the mass of the Pluto-Charon binary
 from Table 10 by \citet{brozovic15} and similarly for $GM_C$.
The semi-major axis of Charon, $a_C$, is also from their Table 9.
 }}
\end{table}

In Table \ref{tab:sats} we list properties of the the 4 minor satellites of the Pluto-Charon
system based on measurements by \citet{showalter15} and improved
measurements by  \citet{weaver16}.
The radii of the equivalent volume spheres, $R_v$ are  listed, and these
we compute using the body axis diameters from Table 2 by \citet{weaver16}.
The spin angular rotation rates, $w$, are computed from the spin periods
and are given in units of  $t_g$ for a density of 1 g~cm$^{-3}$ (see equation \ref{eqn:tg}).
If the actual satellite densities are higher then the spins in dimensionless units would be lower.
The time $t_g^{-1}$ is equal to the angular rotation rate of an orbiting particle just grazing
the surface of a spherical body with radius $R_v$.
In dimensionless units we notice that only Hydra is spinning rapidly 
(at 1/3 of centrifugal breakup).  
compared to a maximum value of approximately 1.
In Table \ref{tab:sats} we also list the ratio of the orbital to spin periods, $P_o/P_s$.
Kerberos and Styx have $P_o/P_s \sim 6$ whereas Nix and Hydra have ratios
of 13.6 and 89.   
All four of the small satellites are spinning much
more rapidly than $w \sim n_o$, or $P_o/P_s \sim 1$.

The masses of the small satellites are not well constrained.  In Table \ref{tab:sats}
the mass ratios $M/M_{PC}$ (satellite mass divided by Pluto-Charon binary mass) are
computed using preferred (or boldface) values from Table 1 by \citet{showalter15}.
These are masses within 1 standard deviation of their dynamical mass constraints
(based on their orbit integrations).

\begin{table*}
\vbox to135mm{\vfil
\caption{\large  Pluto and Charon's minor satellites  \label{tab:sats}}
\begin{tabular}{@{}lllllll}
\hline
                       &    Styx       &  Nix  &  Kerberos &  Hydra \\
\hline
Size (km)        &    $16 \times 9 \times 8$     &   $50\times 35 \times 33$  
          & $19\times 10\times 9$ & $ 65 \times 45 \times 25$ \\
Body axis ratio $b/a$      & 0.56 & 0.70 & 0.53 & 0.69 \\
Body axis ratio  $c/a$      & 0.50  & 0.66 & 0.47 & 0.38 \\
Volumetric Radius $R_v$(km)       &   5.2      & 19.3   & 6.0 & 20.9 \\
Orbital period $P_o$ (days) & $20.16155 \pm 0.00027$ & $24.85463 \pm 0.00003$  & $32.16756 \pm 0.00014$ & $38.20177 \pm 0.00003$ \\
Orbital Semi-major axis $a_o$ (km) &  42,656 &  48,694 & 57,783 & 64,738 \\ 
Spin period $P_s$ (days) & $3.24 \pm 0.07$ & $1.829 \pm 0.009$  & $5.31 \pm 0.10$ & $0.4295 \pm 0.0008$ \\
Period ratio $P_o/P_s$  & 6.2 & 13.6 & 6.06 & 88.8 \\
Period ratio $P_{C}/P_s$ & 1.97 & 3.49 & 1.20 & 14.9 \\
Period ratio $P_o/P_{C}$ &  3.1566 & 3.8913 & 5.0363 & 5.9810 \\
Obliquity $\epsilon_B$ (deg) & 91 & 123 & 96 &  110 \\ 
Spin $w$    &  0.0424 & 0.0752 & 0.026 & 0.320\\
Mass ratio $M/M_{PC}$ & $1\times 10^{-7}$ & $3\times 10^{-6}$ 
   & $8\times 10^{-7}$ & $4 \times 10^{-6}$ \\
$e_g$ (GPa) & $3.2 \times 10^{-5}$&$4.5 \times 10^{-4}$ & $4.3 \times 10^{-5}$&$5.2 \times 10^{-4}$ \\
Love number $k_2$ & $3 \times 10^{-7}$ & $4 \times 10^{-6}$ & $4 \times 10^{-7}$ &
$5 \times 10^{-6}$ \\
$\log_{10} t_{despin}$ (yr)  & 12.7 & 11.6 & 14.4 & 12.2 \\
$\log_{10} t_{wobble}$ (yr)  & 10.5 & 8.0 & 12.2 & 5.4\\
Asphericity $\alpha$ & 1.25 & 1.01 & 1.30 & 1.03 \\
Oblateness parameter $q_{eff}$ & 0.31  &  0.21 &  0.33 & 0.40 \\
Binary quad ratio $(\mu_{PC}/M_{PC}) (a_B/a_o)^2$  &  0.0204 &  0.0157 & 0.0111 & 0.0089 \\
\hline
\end{tabular}
{\\  
The body sizes are diameters $2a,2b,2c$ where $a,b,c$ are body semi-major axes.  Sizes,
orbital periods and  spin rotation periods are from Table 2 by \citet{weaver16}.
The radii of the equivalent volume sphere are computed from the body semi-major axes
as $R_v = (abc)^{1/3}$.
Here satellite obliquity, $\epsilon_B$, is given with respect to Pluto/Charon's north (spin and orbital axes), 
have errors of about $10^\circ$ and are those measured by \citet{weaver16}.
The spin angular rotation rates $w$ are computed from the spin periods
and are given in units of  $t_g$ for a density of 1 g/cc (see equation \ref{eqn:tg}).
The mass ratio $M/M_{PC}$ is given in units of the total mass of the Pluto-Charon binary
and computed using preferred values from Table 1 by \citet{showalter15}.
 The ratio of the satellite orbital period to the period of the Pluto-Charon binary
$P_o/P_C$ is computed using the orbital rotation periods listed here for the minor satellites
and the orbital period for Charon in Table 9 by \citet{brozovic15}.  
The orbital semi-major axes are those listed in Table 2 by \citet{weaver16}.
Energy densities, $e_g$, are computed using equation \ref{eqn:eg} and assuming
a density of 1g/cc.
Love numbers are computed using equation \ref{eqn:k2b},  the shear modulus
of ice $\mu*=4$ GPa, the volumetric radii  listed  here and assuming a density of 1 g/cc.
Tidal spin down times are computed using equation \ref{eqn:tlock}, 
$Q = 100$, and the volumetric radii, spin periods, orbital semi-major axes
and $k_2$ values listed here.
Asphericity, $\alpha$  and oblateness parameters $q_{eff}$  
are computed using equations \ref{eqn:aspher}, and \ref{eqn:qeff}
 and the body axis ratios listed in the table.
 The ratio $(\mu_{PC}/M_{PC}) (a_B/a_o)^2$ is computed using the semi-major axes listed
 here and the masses and semi-major axis listed in Table \ref{tab:PC}.
}
}
\end{table*}

For the minor satellites, we  compute values for energy density $e_g$, Love number $k_2$
and tidal spin-down times $t_{despin}$ and list them in Table \ref{tab:sats}.
The energy density $e_g$ is computed using equation \ref{eqn:eg} (but with $R_v$ replacing $R$), 
the volumetric radii $R_v$ listed in Table \ref{tab:sats} and assuming a density of 1~g/cc.  
 We computed the Love number $k_2$ using our values for $e_g$,
a shear modulus for ice of $\mu* = 4$ GPa and equation \ref{eqn:k2b}.

Tidal spin down times $t_{despin}$ are computed using equation \ref{eqn:tlock}, a tidal dissipation factor
$Q = 100$, the $k_2$,  orbital periods  and mass ratios listed in the Table.
The estimated spin down times (see bottom of Table \ref{tab:sats}) 
exceed the age of the Solar system  and
imply that the satellite spins should not have significantly decreased  due to
tidal dissipation.   This contrasts with 
 the expectation by \citet{showalter15,correia15}  that the minor satellites
of Pluto would be spinning slowly enough to be chaotically
 tumbling. 
The high spin values found by \citet{weaver16} led them to conclude that
 tidal spin down has not yet taken place 
 and the spin down time estimates we have computed 
here support this conclusion (also see \citealt{hastings16}).

The actual spin-down times would be shorter than those listed in
Table \ref{tab:sats} if $Q<100$, corresponding to higher dissipation.
The times would be longer for higher density bodies; $\rho > 1$g/cc and for
stronger bodies. 
The spin down times are estimated for spheres but the torque would only be about twice as large for bodies with the axis ratios of these satellites
(see  \citealt{quillen16b}). 
During a previous epoch of 
higher orbital eccentricity  the torque might have been higher. 
The spin-down time scale estimate also neglects
perturbations by the Pluto-Charon binary and assumes that the spin rate starts
near the maximum value.   The current spin values are 23, 13,  39 and 3 times slower than 
 $t_g^{-1}$ for Styx, Nix, Kerberos and Hydra, respectively.
 The satellite with the shortest spin down time is Nix at $t_{despin} \sim 10^{11.6}$ years
 exceeding the age of the Solar system, even if we divide this by 10 to take into account
 that the spin may have originally been 10 times lower than $t_g^{-1}$.
 Our estimated spin down time is for a spinning body at zero obliquity. 
 Our simulations (not presented here but similar to those presented
 by \citealt{quillen16b}) show 
 that spin-down times are longer at higher obliquity (approximately an order of
 magnitude higher for a prolate body with axis ratio 0.5 at obliquity $90^\circ$)
  so if the body spends much time
 at high obliquity it would not have spun down as far.  
 
\subsection{Wobble decay times}

It is common to assume that bodies are spinning about a principal axis because
the timescale for wobble to decay is shorter than the spin-down time 
(see \citealt{burns73,peale77,harris94,frouard16}). 
Due to tidal dissipation
the wobble decays approximately on an  timescale 
\begin{equation}
t_{wobble} \approx \frac{3 G C Q}{w_{init}^3 k_2 R^5} \label{eqn:twobble}
\end{equation}
\citep{peale77}
where $C$ is the maximum moment of inertia about a principal body axis.
The ratio of the wobble decay to spin-down time 
\begin{equation}
t_{wobble}/t_{despin} \approx   9(n_o/w_{init})^4  \label{eqn:twob}
\end{equation}  
(the ratio of equation \ref{eqn:twobble} and \ref{eqn:tdespin}; \citealt{gladman96}). 
We compute $t_{wobble}/t_{despin}$  from the ratio of current spin and orbital periods
finding $t_{wobble}/t_{despin}  \sim  6 \times 10^{-3}$ for Styx, and Kerberos, 
$3 \times 10^{-4}$ for Nix and  $10^{-7}$ for Hydra.    The wobble decay
timescales, computed
using the breakup angular rotation rate,  are also listed Table \ref{tab:sats}.
Hydra and  Nix are spinning fast enough that their wobble decay times are much
shorter than the age of the Solar system.
Taking spins near their current
values (rather than a near breakup value), reduces the spin-down time by 20
for Styx and Kerberos.   This reduces the wobble decay time for Styx to near the
age of the Solar system, but not Kerberos.  Kerberos could be wobbling. 
Keeping in mind that  equation \ref{eqn:twobble} is approximate,
 only Nix and Hydra are likely to be spinning
about a principal body axis.
If tumbling in one of the minor satellites was excited by an encounter (collision or tidal encounter)
or a spin resonance at some time well after formation, the long wobble decay
timescales imply that the body could still be tumbling today  (and this is particularly relevant
for Kerberos).

\subsection{Asphericity}

The width of spin-orbit resonances depends on an asphericity parameter
\begin{equation}
\alpha \equiv \sqrt{3 (B-A)/C}  \label{eqn:alpha} 
\end{equation}
\citep{wisdom84}
where $A < B \le C$ are the three moments of inertia (eigenvalues of the moment of inertia matrix).
The frequency 
$ \omega_{so} = \alpha n_o$ 
 is the frequency of small-amplitude oscillations (librations) of a satellite in synchronous resonance 
 (see  \citealt{wisdom84}). 
For a homogeneous triaxial ellipsoid with body axes $a> b>c $,
the moments of inertia are $C = \frac{M}{5}(a^2 + b^2)$, 
$B = \frac{M}{5}(a^2 + c^2)$, and
$A= \frac{M}{5}(b^2 + c^2)$ 
giving 
\begin{equation}
\alpha = \sqrt{ \frac{3(a^2 - b^2)}{a^2 + b^2}} = \sqrt{\frac{3(1 - (b/a)^2)}{1 + (b/a)^2}}.
\label{eqn:aspher}
\end{equation}
This only depends on the body axis ratios in the orbital plane 
(assuming a body with parallel spin and orbital
normal and spinning about a principal axis). 
An oblate body has $\alpha=0$.
We compute asphericities for Pluto's minor satellites from the body axis ratios by \citet{weaver16}, 
and they are listed in Table \ref{tab:sats}.  The minor satellites are all sufficiently elongated that $\alpha \ga 1$.

\subsection{Spin Precession Frequency}

The spin axis of a non-spherical body spinning about its  principal axis
in orbit about a central mass precesses.
Using equation 1  by \citet{gladman96} 
the tidally induced   precession rate 
\begin{equation}
\dot \Omega_s \approx - \frac{S}{w} \cos \epsilon_o,
\end{equation}
where $\epsilon_o$ is the obliquity (angle between spin axis and orbit normal) and
\begin{equation}
S \equiv \frac{3}{2} n_o^2 \frac{C - (A +B)/2}{C}.
\end{equation}
 When the body is spinning about a principal body axis,
 the angles $\Omega_s, \epsilon_o$ are Euler angles and the precession rate $\dot \Omega_s$
 is the rate the spin vector precesses about the orbit normal.
It is convenient to define a parameter related to an
effective oblateness of the body averaged over its spin (when spinning
about the maximum principal axis) 
\begin{equation}
q_{eff} \equiv  \frac{C - (A +B)/2}{C} = 
\frac{(1 + (b/a)^2)/2 - (c/a)^2}{1 + (b/a)^2}  \label{eqn:qeff}
\end{equation}
where we have used ellipsoidal body semi-major axis ratios $a,b,c$.
The parameter $q_{eff}$ for each of Pluto and Charon's minor satellites are also listed
in Table \ref{tab:sats} and range from 0.2 (Nix) to 0.4 (Hydra).
Using $S$ and $q_{eff}$ the body spin precession rate divided by the orbit mean motion 
\begin{equation}
\frac{\dot \Omega_s}{n_o}  \approx - \frac{3}{2} q_{eff} \frac{n_o}{w} \cos \epsilon_o 
\approx - \frac{3}{2} q_{eff} \frac{P_s}{P_o} \cos \epsilon_o \label{eqn:spin_prec}
\end{equation}
and on the right we have written the precession rate
in terms of the ratio of spin and orbit periods where 
the spin period $P_s = 2\pi/w$ and the orbital period $P_o \approx 2\pi/n_o$.
At low obliquity and for $q_{eff} = 0.3$, the precession rate is rapid,
at $\frac{\dot \Omega_s}{n_o}  \approx \frac{1}{2} \frac{P_s}{P_o}$.
For a spinning body orbiting about a binary rather than  a point mass, equation \ref{eqn:spin_prec}
likely underestimates the precession rate by a factor that depends on the quadrupole
moment of the averaged binary gravitational potential.  This moment is 
given in the next subsection and is small for Pluto and Charon's satellites 
(only 1--2\% of the monopole) 
so equation \ref{eqn:spin_prec} is a pretty good
estimate.


The spin precession rate is independent of body size or $R_v/a_o$.  
For numerical convenience we can increase the
spin in units of $t_g$ and this is equivalent to reducing the
density of the simulated spinning satellite.
As long as we maintain the period ratios (orbital to binary to spin) the secular precession 
frequencies are not modified.

\subsection{Orbit Precession Frequencies}

Resonance location often depends on secular frequencies such as the precession rate
of the longitude of the ascending node.  
The gravitational potential of the binary averaged over its orbit is similar to an oblate body. It
has a quadrupole gravitational moment 
giving an effective gravitational harmonic coefficient 
\begin{equation}
J_2^{eff,B} = \frac{1}{2} \frac{\mu_B}{M_B}
\end{equation}
(see problem 6.3 by \citealt{M+D})
where $\mu_B$ is the reduced mass of the binary and $M_B$ is the total mass of the binary.
Inserting this $J_2$ into equations for orbital precession  frequencies  \citep{greenberg81} 
and taking lowest order terms
\begin{eqnarray}
n&\approx & n_o \left[ 1 + \frac{3}{8} \left(\frac{a_B}{a_o}\right)^2 \frac{\mu_B}{M_B} \right]  \label{eqn:n} \\
\dot \Omega_o &=& 
 -n_o \frac{3}{4} \left(\frac{a_B}{a_o}\right)^2 \frac{\mu_B}{M_B}  \label{eqn:dot_Om}
 \\
\dot \varpi_o &=& 
n_o \frac{3}{4} \left(\frac{a_B}{a_o}\right)^2 \frac{\mu_B}{M_B}  \label{eqn:dot_varpi},
\end{eqnarray}
with  \begin{equation}
n_o \equiv  \sqrt{\frac{GM_B}{a_o^3}}. \label{eqn:no}  \end{equation}
Here $\dot \Omega_o$ is the precession rate for the longitude of the ascending node, $\Omega_o$,
and $\dot \varpi_o$ is the apsidal precession rate or the precession for the longitude of pericenter, $\varpi_o$.
We have subscripted orbital parameters and frequencies with an $o$ so as to differentiate them
from spin related quantities, but we also use the subscript to denote
quantities that are based on osculating orbital elements.
Here $n$ is intended to approximate the sidereal mean motion whereas the osculating $n_o$
is dependent on the osculating semi-major axis $a_o$. 
For a particle in the binary orbital plane, the orbital period is computed from  $n$ not $n_o$.
The osculating orbital semi-major axis, $a_o$, is  computed for
a particle in the binary plane assuming
a Keplerian orbit and with respect to the mass and center of mass of the binary
(see discussion by \citealt{greenberg81} and \citealt{renner06}).
As the orbital period and orbital precession frequencies depend on the ratio
$ \frac{\mu_B}{M_B} \left(\frac{a_B}{a_o}\right)^2$ we have computed this for the four minor satellites
using values for the binary in Table \ref{tab:PC} and semi-major axes of the satellites in Table \ref{tab:sats}
and we include the computed values for this ratio at the bottom of Table \ref{tab:sats}.


\section{Simulations of the Spinning Minor Satellites} \label{sec:sims}

We begin by describing our simulation technique.  Two types of simulations are
 carried out, simulations with tidal dissipation alone (discussed in section \ref{sec:nd}) and simulations
with a slowly drifting apart central  binary (discussed in section \ref{sec:bin}).  
In both settings we track the spin and orbital
evolution of a spinning satellite in orbit about a binary that represents Pluto and Charon.

\begin{figure}
\includegraphics[width=3.0in]{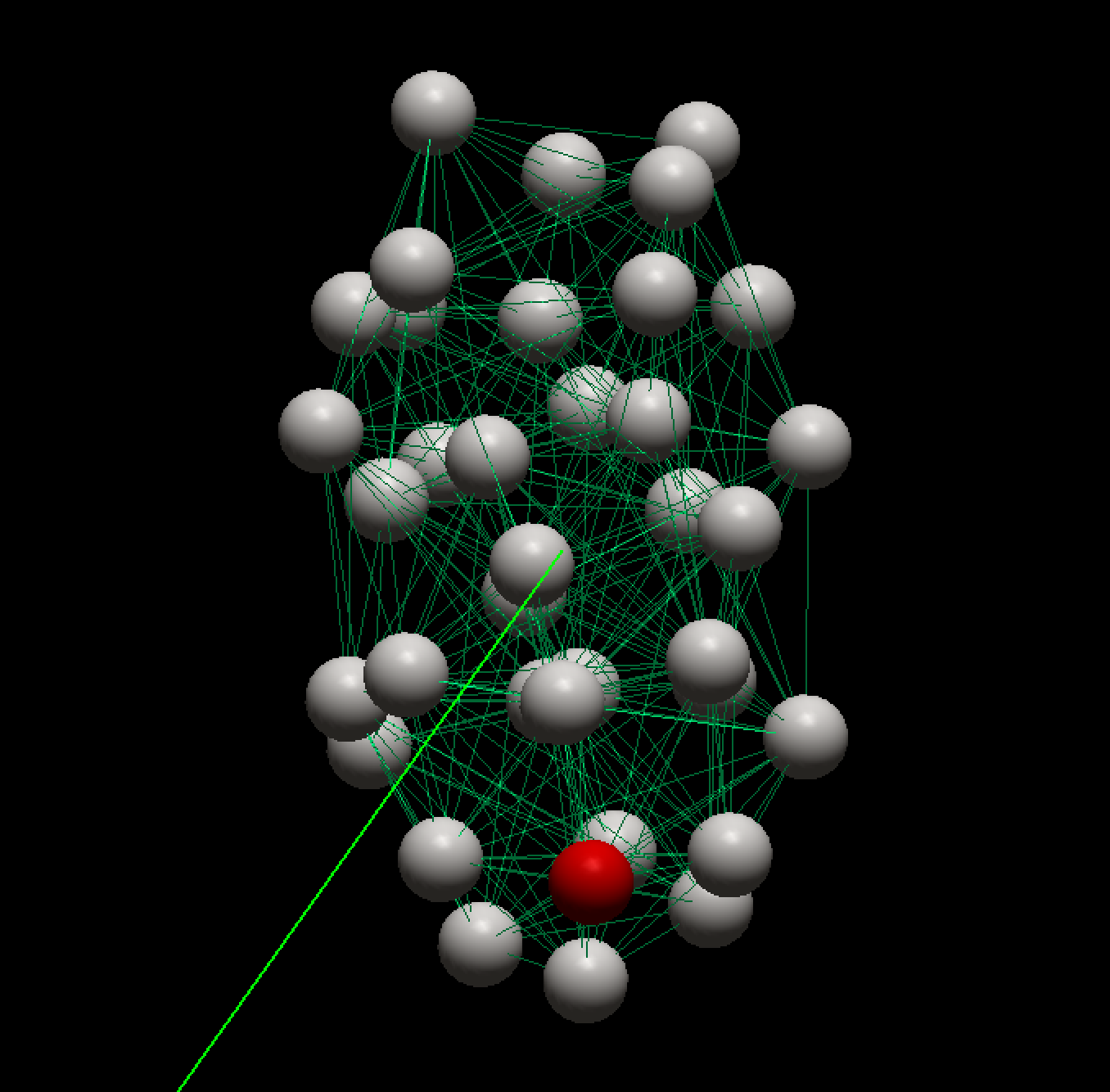}
\caption{A simulation snapshot showing an oblong spinning body.
This snapshot is from a simulation exploring Styx's spin evolution and the body
has axis ratios similar to Styx.  The snapshot shows the spinning body only but it is in orbit
about a binary (see Figure \ref{fig:bin}). The grey spheres show the mass nodes and the green lines
the springs.  One sphere is marked red so the body rotation can be more easily seen. 
   The bright green line connects the center of mass of the spinning body to the primary in the binary.
The body only contains 38 mass nodes and about 350 springs.  The small number of 
particle nodes allows us to run the simulation for many spin and orbital rotation periods.
\label{fig:snap}}
\end{figure}

\subsection{Description of mass-spring model simulations}

To simulate tidal viscoelastic response of non-spherical bodies
 we use a mass-spring model \citep{quillen16a,frouard16,quillen16b}
 that is built on the modular N-body code rebound  \citep{rebound}.  
 Springs between mass nodes are damped and
so the spring network approximates the behavior of a Kelvin-Voigt viscoelastic 
solid with Poisson ratio of 1/4 \citep{kot15}.
\citet{frouard16,quillen16b} considered a binary in a circular orbit.  
One of the masses was a spinning body resolved with masses and springs. 
The other body (the tidal perturber) was modeled as a point mass.
Here we consider three bodies, a spinning body resolved with masses and spring in 
orbit about a binary comprised
of two point masses (representing Pluto and Charon); see Figure \ref{fig:bin}.  
The total binary mass is $M_B$ and the ratio
of the smaller to larger mass in the binary, $q_B$.   For our simulations we set the mass ratio
$q_B=0.12$ to be equal to the Charon to Pluto mass ratio.  The total binary mass
is set to $10^6$ in units of $M$ and is about the right order of magnitude for the
ratio of a minor satellite to the sum of Pluto and Charon's mass, though Nix and Hydra are more
massive than Kerberos and all three more massive than Styx  (see Table \ref{tab:sats} for mass ratios).
Each simulation only tracks three masses, the binary and the resolved spinning body, so
 our simulations neglect dynamical interactions between the minor satellites themselves
  and tidal interaction between Pluto and Charon.  

The mass particles in the resolved spinning body are subjected to three types of forces: the gravitational forces acting on every pair of particles in the body and with two massive point mass companions, and the elastic and damping spring forces acting only between sufficiently close particle pairs. Springs have a spring constant $k_s$ and a damping rate parameter $\gamma_s$.  
When a large number of particles is used to resolve the spinning body the mass-spring model behaves
like a continuum solid.
The number density of springs, spring constants and spring lengths set the shear modulus, $\mu$, whereas
the spring damping rate,
$\gamma_s$, allows one to adjust the shear viscosity, $\eta$, 
and viscoelastic relaxation time, $\tau_{relax} = \eta/\mu$. 
The tidal frequency in units of the relaxation time $\bar \chi = | \sigma_t| \tau_{relax}$ (see section 2.3  by \citealt{frouard16}) and for $\bar\chi <1$, the quality function for our mass-spring model has
$k_2(\sigma_t) \epsilon_2(\sigma_t) \approx k_2 \bar \chi$.    The attenuation properties
of cold icy bodies is dependent on composition, porosity, temperature and frequency
(\citealt{castillorogez12,ice13}) and so is not well constrained.  
 In our simulations we chose
$\gamma$ so as to remain in the linear regime
where the quality function is proportional to $\bar\chi$ (and approximately giving a constant time lag 
tidal dissipation model).   

Our previous studies \citep{frouard16,quillen16b} were restricted to bodies spinning about
their principal axes and with parallel spin and orbital axes. 
Here we allow the spinning object
to have an initial non-zero obliquity.  We measure the tilt of the spinning body
in two ways.    The obliquity $\epsilon_o$  is the angle between 
the spin angular momentum vector of the resolved body and its orbit normal (the direction
of orbital angular momentum). The orbit normal and orbital elements for the spinning body such as
its inclination, eccentricity and semi-major axis, $a_o$, are
measured with respect to 
 the center of mass of the Pluto-Charon binary (and using the vector between
 the center of mass of the spinning body and the center of mass of the binary).
The obliquity $\epsilon_B$ is the angle between the spin angular momentum
vector of the resolved body and the binary orbit normal.  
The binary's mean motion, $n_B$,
and semi-major axis, $a_B$, are computed neglecting the much lower mass
spinning body.  The orbital semi-major axis is computed
using the total mass of the binary and  coordinates measured
from the center of mass of the binary.  The two sets of coordinates are essentially Jacobi coordinates
(see Figure \ref{fig:bin}).

In our previous studies,
we measured instantaneous tidal torques and so we resolved the spinning
body with numerous particles. 
Here we aim to explore longer timescale behavior. Instead of maintaining or increasing the number
of particles in the resolved body, we decrease it.    Rather than a thousand or more particles
in the resolved body we typically have only 40.  The particle number does not
change during a simulation, however different simulations
have slightly different numbers of particles 
as particle positions are initially randomly generated.
 A simulation snap shot is shown in Figure \ref{fig:snap}.
The small number of particles or mass nodes allows us to run for
many thousands of spin rotation periods. 
\citet{frouard16} measured a 30\% difference between the drift rate computed from the simulations
and that computed analytically.   We do not try to resolve this discrepancy here but instead 
study the long timescale evolution of spin and obliquity with the goal of trying to
understand processes that would have affected minor satellite obliquity. 

We work with mass in units of $M$,  the mass of the spinning body,
distances in units of volumetric radius, $R_v$, the radius of a spherical body with the same
volume, time in units of $t_g$  (equation \ref{eqn:tg}) and
elastic modulus $E$ in units of $e_g $ (equation \ref{eqn:eg})
which scales with the 
gravitational energy density or central pressure.  
Initial node distribution and spring network are those of the triaxial ellipsoid random spring model
described by \citet{quillen16b}.  
 For the random spring model, particle positions are drawn from an isotropic uniform distribution but only accepted into the spring network if they are within the surface bounding a triaxial ellipsoid, 
 $x^2/a^2 + y^2/b^2 + z^2/c^2 = 1$, and if they are more distant than $d_I$ from every other previously generated particle. Here $a,b,c$ are the body's semi-major axes.
Once the particle positions have been generated, every pair of particles within $d_s$ of each 
other are connected with a single spring.    
Springs are initiated at their rest lengths.  
The body is initially a triaxial ellipsoid and if it were not rotating it would remain a triaxial ellipsoid.
For the mass spring model,
Young's modulus,  $E$, is computed 
 using equation 29 by  \citet{frouard16}  and using
the entire triaxial ellipsoid volume (bounded by $x^2/a^2 + y^2/b^2 + z^2/c^2 = 1$), however with
only 40 or so particles this value is approximate.  With only about 40 particles
and  about 400 springs, the spinning body does not well approximate a continuum solid with its material properties.  However our simulation technique can show spin-orbit resonance capture
and tumbling and it 
accurately computes time dependent torques arising from the Pluto-Charon binary.

A fairly low value of Young's modulus (in units of $e_g$) was used so that the body was soft, 
reducing the integration time required to see tidal drifts in orbital semi-major axis, spin and obliquity due
to dissipation in the springs.
The viscoelastic relaxation timescale and associated
tidal frequency $\bar \chi$ are computed  as done previously  \citep{frouard16,quillen16b}.
%
The orbital periods are about 100 times larger than $t_g$, however our time-step $dt$ is set by the time
it takes elastic waves to travel
between particles in the resolved body and so are much smaller than
the orbital period.
Each simulation required a few hours of computation time on a 2.4 GHz Intel Core 2 Duo from 2010.

\begin{table}
\vbox to70mm{\vfil
\caption{\large  Common simulation parameters  \label{tab:common}}  
\begin{tabular}{@{}lllllll}
\hline
Binary mass      & $M_B$        & $10^6$  \\
Time step           & $dt$   & 0.004  \\
Simulation outputs & $t_{print}$ & 100 \\
Total integration time  & $T_{int}$ &  1 --$ 3 \times 10^6$ \\
Minimum particle spacing & $d_I$ & 0.47 \\  
Maximum spring length & $d_s$ & $ 2.55 d_I$ \\ 
Spring constant   & $k_s$              & 0.05  \\ 
Number of particles  & $N$ &  $\sim 40$ \\
Number of springs   & $N_S$ & $\sim 375$ \\
%
Young's modulus & $E/e_g$      & $\sim 0.6$   \\
\hline
\end{tabular}
{\\  Notes.   Mass and volume of the spinning body are the same for all simulations.
Distances  $d_s$ and $d_I$ and spring constant $k_2$ 
are used to generate the random spring network (see \citealt{quillen16b}).
$N$ and $N_S$ refer to the number of particle nodes and springs in the resolved spinning body and
vary by a few between simulations as the initial particle positions are generated randomly.
Points in our subsequent figures are separated in time by $t_{print}$.
}
}
\end{table}

\begin{table} 
\vbox to65mm{\vfil
\caption{\large  Simulation parameters for each simulated satellite \label{tab:series_body}}
\begin{tabular}{@{}llllllll}
\hline
Simulated satellite & & Styx &  Nix & Kerberos \\
\hline
Body axis ratio    &   $b/a$       & 0.56 & 0.70 & 0.53  \\
Body axis ratio    &   $c/a$       & 0.47 & 0.67 &0.47   \\
initial spin  &  $w_{init}$  & 0.5  &  0.72 &  0.5 \\
Initial orbital semi-major axis & $a_o$ & 600 & 830 & 600\\
Initial mean motion & $n_o$ & 0.068 & 0.042 & 0.068 \\
Initial orbital period & $P_{o,init}$ &  93.0 &151.1 & 92.7 \\
\hline
\end{tabular}
{\\  The orbital period was computed using equation \ref{eqn:n} and takes into account
the quadrupole moment of the binary. The mean motion is computed without correction
and is based solely on the osculating semi-major axis $a_o$ (equation \ref{eqn:no}). 
Orbital period is in units of $t_g$ (equation \ref{eqn:tg}), 
mean motion in units of $t_g^{-1}$ and
semi-major axis is in units of $R_v$, the radius of the equivalent volume sphere. 
}
}
\end{table}

\begin{table*}
\vbox to50mm{\vfil
\caption{\large  Parameters for simulations with tidal dissipation alone\label{tab:series_nd}}
\begin{tabular}{@{}lllllllllll}
\hline
Simulation name & & Styx-t1 & Styx-t2 & Styx-t3 & Styx-t4 & Nix-t1 & Nix-t2 & Ker-t1 & Ker-t2\\
\hline
Spring damping rate   & $\gamma_s$ & 0.01 & 0.01 & 0.01 &0.01 & 0.1 &0.1 & 0.01 & 0.1 \\
Tidal frequency          & $\bar \chi$   &0.002 & 0.002 & 0.002 &0.002 & 0.04 &0.04 & 0.002 & 0.02 \\
Initial binary semi-major axis & $a_B$ & 290 & 290 & 185 &305 & 340 & 326 & 206 & 206\\
Binary mass ratio    & $q_B$               & 0.12  & $10^{-9}$ & 1 &0.12 & 0.12 & 0.12 &  0.12 & 0.12 \\
\hline
\end{tabular}
{\\  Initial obliquities for these simulations is $\epsilon_{init} = 20^\circ$.
}
}
\end{table*}

\begin{figure*}
\includegraphics[width=6.5in]{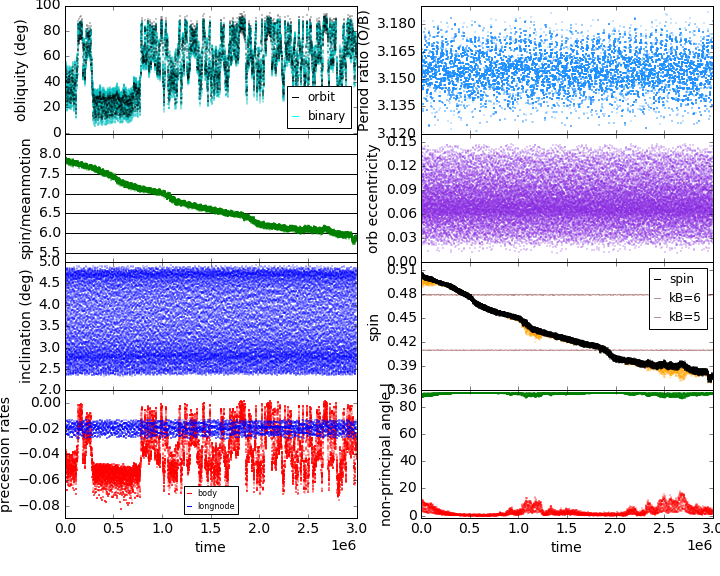}
\caption{Tidal evolution for Styx and showing the Styx-t1 simulation with
 parameters listed in Tables \ref{tab:common}, \ref{tab:series_body} and \ref{tab:series_nd}.
Each sub-panel shows evolution of a quantity. The horizontal axes are the same and in units of time $t_g$
or about 2000 s. 
The orbital period of the spinning body is listed in Table \ref{tab:series_body} 
and is about 90.
The top left panel shows obliquity evolution of the spinning body.  The black dots show obliquity 
measured with respect to the body's orbit normal. The blue dots show obliquity measured with
respect to the orbit normal of the binary.
The second from top left panel  shows the body spin divided by the orbital mean motion with
black horizontal lines giving the location of spin-orbit resonances.
The third-left panel shows orbit inclination.
The bottom left panel shows the spin precession rate (in red) and the precession
rate of the longitude of the ascending node of the orbit (in blue) both divided by $n_o$.
The top right panel shows the ratio of the orbital period (for the spinning body) divided
by that of the binary.  The second panel (from top) on the right shows orbital eccentricity.
The third right panel shows body spin (in black) and  the brown lines  on this plot shows the location
of the nearest spin-binary resonances in the form of equation \ref{eqn:kB}
and labelled with their integer $k_B$.  
Also plotted in this panel  (in orange) is the component of the spin vector in the direction
of the maximum principal axis.  This only differs from the spin when the body is tumbling.
The bottom right panel shows in red the non-principal angle $J$ (the
angle between spin angular momentum and the axis of the body's maximum principal
axis of inertia) and in 
green the angle between spin angular momentum and the minimum principal body axis.  Both angles
are in degrees and they
only differ from 0 or $90^\circ$ when the body is tumbling.
This simulation illustrates intermittent obliquity variations, episodes of tumbling and jumps in spin associated with crossing spin-binary resonances with indices $k_B = 5$ and $6$ (and $k_B=5$ 
is the lower one). 
\label{fig:styx-nd}}
\end{figure*}  

\begin{figure*}
\includegraphics[width=6.5in]{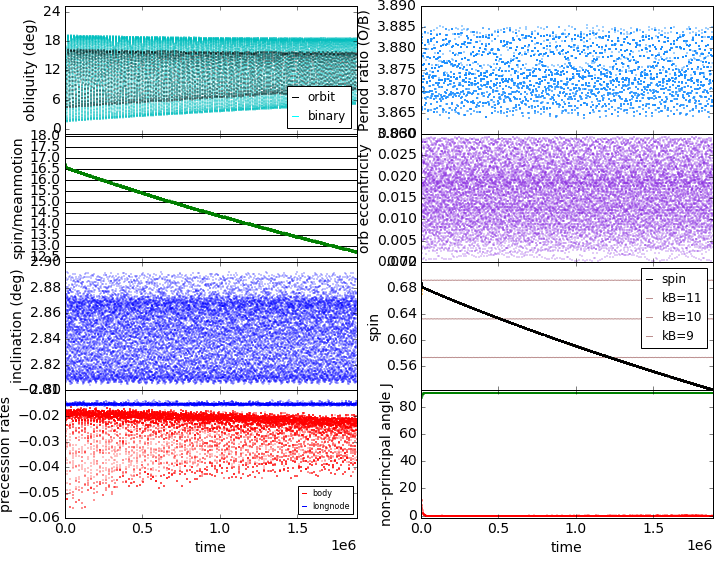}
\caption{Tidal evolution for Nix and showing  the Nix-t1 simulation.
Similar to Figure \ref{fig:styx-nd} and with parameters listed in Tables \ref{tab:common} -- 
 \ref{tab:series_nd}.  This simulation illustrates that at the spin rate of Nix, spin-orbit resonances
 and spin-binary resonances
do not affect the spin or obliquity of Nix.
\label{fig:nix-nd}}
\end{figure*} 

\begin{figure*}
\includegraphics[width=4.9in]{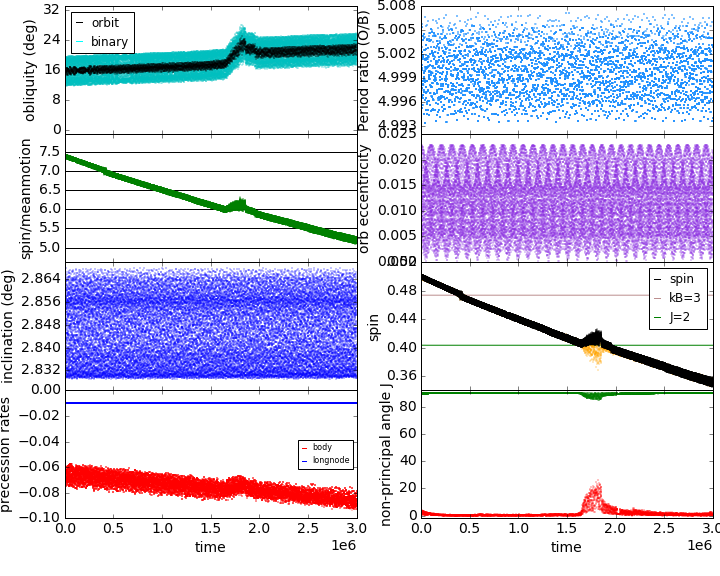}
\includegraphics[width=4.9in]{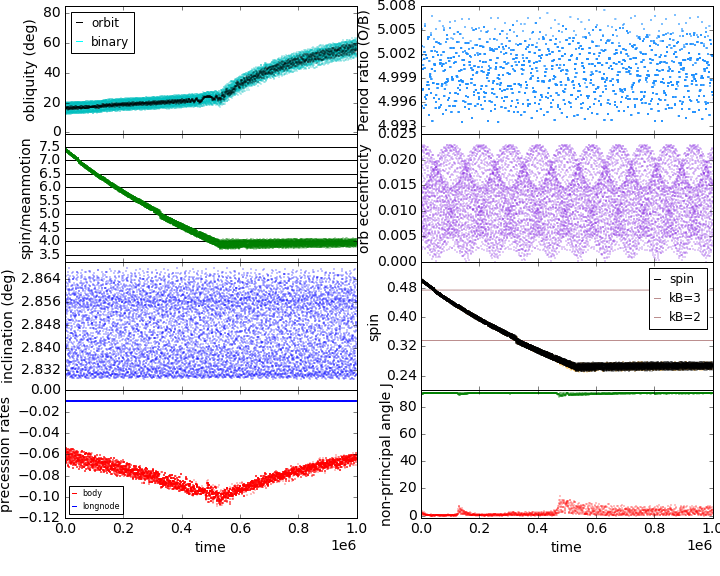}
\caption{Tidal evolution for Kerberos and showing a)  the Ker-t1 and b) the Ker-t2 simulations.
These simulations differ only in the level of their tidal dissipation.
Similar to Figure \ref{fig:styx-nd} and with parameters listed in Tables \ref{tab:common} -  \ref{tab:series_nd}.
The Ker-t1 simulation (figure a) illustrates a temporary spin resonance capture 
that also causes some tumbling.
Both simulations illustrate  jumps in spin from crossing  spin-binary resonances.
The  resonance near $w/n_o \sim 4$ that causes the obliquity to rise is
 at a spin value well below that of Kerberos's current value. 
\label{fig:ker_nd}}
\end{figure*} 

\subsubsection{Description of Simulation Figures}

We begin with simulations intended to be near current orbital  period and spin ratios and with
spinning body exhibiting tidal spin-down due to dissipation in our simulated springs.
Common simulation parameters are listed in Table \ref{tab:common}.  Our simulations are labelled
according to which body is simulated and parameters that depend upon which body
is simulated are listed in Table \ref{tab:series_body}.
The  different body semi-major axis ratios, appropriate for Styx, Nix and Kerberos are
the same as in Table \ref{tab:sats}.  
The simulations have different initial spins and semi-major axes.
The spin values are chosen so the initial spin to orbit periods are similar to (but slightly higher than)
the  values listed in Table \ref{tab:sats} and measured by \citet{weaver16}.
As time is given in units of $t_g$,  Table \ref{tab:series_body} also lists the
initial orbital period of the minor satellite (the resolved spinning body) about
the center of mass of the binary.  

The initial conditions are chosen to have ratios of spin to orbital period, orbital to binary period
and spin precession to orbital period similar to Pluto and Charon's minor satellites.
However,
we set the orbital semi-major axes, in units of minor satellite radius $R_v$, smaller than
the actual ratios so as to decrease the timescale required for tidal evolution to take
place and allow us to see  long timescale phenomena associated with spin-down in the simulations.
This means that the spin rates in units of $t_g^{-1}$ are larger than
their actual values though we approximately maintain the ratio of spin and orbital period
and ratio of orbit and binary period.   Increasing the spin (in units of $t_g^{-1}$) 
is equivalent to simulating
a larger lower density body  as $t_g$ depends on density.
Because $w/n_o$ is larger for Nix than for Kerberos and Styx, the Nix simulations
required larger initial semi-major axis  (in units of $R_v$) 
so as to keep the ratios $w/n_o$ and $n_o/n_B$
similar to the actual values for this satellite.   We don't simulate Hydra because
 of its high actual spin value.   To maintain its spin to orbit period rate and
 keep its spin below breakup we would require a large semi-major axis and much
 longer numerical integration times.

Parameters that differ for simulations with tidal dissipation alone are listed in Table \ref{tab:series_nd}.
Dissipation is set by the spring damping parameter, $\gamma_s$ and adjusted
so that the tidal frequency $\bar\chi$ (listed in Table \ref{tab:series_nd}) is less than 1, so as
to remain in a linear regime where the quality function is proportional to $\bar \chi$ 
(a constant time lag regime, but not constant dissipation $Q$ or constant phase lag regime). 
The Styx simulations have initial orbital period to binary period ratio
about 3, that for Nix  about 4 and those for Kerberos about 5.

In Figure \ref{fig:styx-nd} we plot quantities as a function of time  measured from the outputs of
a simulation for Styx.   Points are plotted each simulation output, ($t_{print}$ in Table \ref{tab:common}).
The horizontal axes for each panel are the same
and in units of  $t_g$.  
Orbital elements for the spinning satellite 
are computed using the initial orientation of the binary as a fixed reference frame (with the binary orbital
plane defining zero inclination) and using 
the mass of the binary.  We compute the osculating orbital elements at each simulation output using
the center of mass position and velocity of the spinning satellite measured with respect
to the center of mass of the binary. 
Orbital inclination in degrees is shown in the third (from top)-left panel, and orbital
eccentricity is shown on the second (from top)-right panel of the figure.
The top right panel shows the ratio of the orbital period (for the spinning body) divided
by that of the binary.   

The angular momentum of the spinning body, ${\bf L}_s$, is computed at each simulation output by summing
the angular momentum of each particle node, using node positions and velocities
measured with respect to the center of
mass of the body.  The moment of inertia matrix of the spinning body, in the fixed reference frame,
$\bf I$, is similarly computed.
The instantaneous spin vector $\bf w$ is computed by multiplying the spin angular momentum
vector by the inverse of the moment of inertial matrix, ${\bf w} = {\bf I}^{-1} {\bf L}_s$.  
The spin $w = |{\bf w}|$ is the magnitude of this spin 
vector and shown divided by the osculating orbital mean motion, $n_o$, in the second (from top)-left panel.
Black horizontal lines show the location of spin-orbit resonances on the same panel.

\citet{correia15} identified spin-binary resonances at spins
\begin{equation}
w  = n_o + \frac{k_B}{2} (n_B - n_o) \label{eqn:kB}
\end{equation}
for each non-zero integer $k_B$.
 The spin itself is shown in the third (from top)-right panel (in units of $t_g^{-1}$) 
 along with spin-binary resonances, plotted in brown, and labelled with integer $k_B$. 

The dot product of the spinning body's spin angular momentum vector and current
orbit normal gives $\cos \epsilon_o$ where $\epsilon_o$ is an obliquity with respect to
the orbit normal.
This is shown with black dots in the top, left panel.
The obliquity computed from the dot product of the  spin angular momentum vector and 
binary orbit normal,  $\epsilon_B$, is plotted with blue dots on the same panel.
A comparison of the two obliquities can determine whether the object
is in a Cassini state where the body's spin precession rate matches that of the longitude
of the ascending node, $\dot \Omega_s \sim \dot \Omega_o$.

At each simulation output we diagonalize the moment of inertia
matrix and identify the eigenvectors (directions) associated with each principal body axis.
The angle between the body's spin angular momentum and the direction of
the maximum principal axis in degrees is shown with red dots on the bottom right panel.
This angle, $J$, is sometimes called the non-principal angle and it is one of
the Andoyer-Deprit variables \citep{celletti}.  The angle between the 
spin angular momentum and the direction of the
the minimum principal axis is plotted on the same panel with green dots.
When the body is spinning about the principal axis $J=0$ and the other angle
is $90^\circ$.  Only when the body is tumbling do these angles differ from 0 and $90^\circ$.
Also plotted in the third (from top)-right panel (in orange) is the component of the spin vector in the direction
of the maximum principal axis.  Only when the satellite is spinning about the maximum principal axis is
this equal to the spin.

At each simulation output we compute the osculating orbital longitude of the ascending node, $\Omega_o$.
The time derivative of this (computed from the simulation output differences)
is used to compute instantaneous measurements of the orbit precession rate
$\dot \Omega_o$. The precession rate divided by the initial mean motion $\dot \Omega_o/n_{o,init}$  is plotted with blue dots in the bottom left panel where $n_{o,init}$ is the osculating mean motion at the beginning of the simulation.
At each simulation output 
the body spin angular momentum is projected into the initial orbital plane of the binary
and the time derivative of the angle in this plane used to compute the spin precession rate $\dot \Omega_s$.
The angle $\Omega_s$ is that between the x-axis and the spin angular momentum ${\bf L}_s$ vector after 
projection into the binary orbital plane, 
$\Omega_s = {\rm atan2} (({\bf L}_s \cdot \hat {\bf y}),({\bf L}_s \cdot \hat {\bf x}))$ where $x,y$ coordinates
span the binary's orbital plane.  The spin precession rate
divided by  $n_{o,init}$
is plotted with red dots in the same panel as $\dot \Omega_o$.  When the two precession rates coincide,
the body is in a Cassini state.
The body spin precession rate curve (red line in bottom left panel) usually resembles 
the obliquity trajectory because the body precession rate is sensitive to obliquity,   
 though $|\dot \Omega_s|$ also slowly increases as the body spins down
 (see equation \ref{eqn:spin_prec}).
 
 When the body is spinning about a principal body axis,
 the angles $\Omega_s, \epsilon_B$ are Euler angles in
 a coordinate system defined by the binary orbital plane and using the initial binary orientation
 to give an $x$ coordinate axis  direction.

\section{Tidal spin down alone}  \label{sec:nd}

In Figures \ref{fig:styx-nd} - \ref{fig:ker_nd} we show simulations for tidal evolution (spin-down)
for Styx, Nix and Kerberos.    The spinning bodies are begun at an obliquity near $20^\circ$ and at small
orbital inclinations of a few degrees.  Tidal dissipation, set by the spring damping rate $\gamma_s$, 
 is chosen to be low enough that 
the bodies do not spin down completely during the integration.
Nix has a higher spin with $w/n_o \sim 14$.  Figure \ref{fig:nix-nd}a shows
that Nix crosses spin-orbit resonances without much affect on body spin or obliquity (see second from top
left panel).  Neither is a spin resonance with the binary important 
(see third from top right panel).
Obliquity oscillations arise due to precession of the spinning body with respect 
to the quadrupole potential of the binary, the small but non-zero orbital inclinations and the proximity
to the 4:1 resonance with the binary.  

\subsection{Styx's obliquity intermittency}

With spin rate lower than for Nix, Styx (see Figure \ref{fig:styx-nd}) might be
 affected by passage through spin-orbit or spin-binary resonances.   
 The ratio of spin to mean motion exhibits
kinks (see second from top and left panel) near spin-orbit resonances
and near spin-binary resonances (see third right panel).  The body also
experiences episodes of tumbling
(see bottom two right panels) though these might be due to resonant perturbation exciting
 a tumbling or nutation frequency rather than a spin-orbit or spin-binary resonance or an instability
associated with a secular spin resonance.
 
Obliquity variations are intermittent in the Styx-t1 simulation.
The regime of weak resonance overlap is often associated with intermittent chaotic behavior.
By comparing this simulation to similar ones, we attempt to identify the source of the chaotic behavior.
 We first  checked that a similar simulation but starting at lower obliquity 
 shows same phenomena as the
Styx-t1 simulation.    It does but it takes longer to reach high obliquity.

Does the binary play a role in causing Styx's obliquity intermittency?   
 A simulation with identical parameters but
with an extremely low mass binary (the Styx-t2 simulation with mass ratio $q_B= 10^{-9}$) 
does not show intermittent obliquity variations
and the simulation lacks jumps or kinks 
in the spin decay trajectory as spin-orbit resonances are crossed.   

The binary also affects secular precession frequencies so perhaps
the weak  spin-orbit and spin-binary resonances and strong obliquity variations imply that
multiple secular resonances affecting spin are responsible for Styx's high current obliquity.
To test this possibility we ran the Styx-t3 simulation, similar to the Styx-t1 simulation but
with a higher binary mass ratio $q_B = 1$ and
 a smaller binary semi-major axis so that binary quadrupole moment is the same
as in the Styx-t1 simulation.  The binary is more compact in the Styx-t3 simulation and $n_B$ about
twice that of the Styx-t1 simulation.   This simulation was dull,
lacking chaotic behavior.  Spin-orbit resonances  when crossed did not affect spin or body
orientation, though because the binary period was shorter, the spin-binary resonance with 
$k_B =2$ was noticeable; it caused a jump in spin as it was crossed.  
We conclude that secular perturbations alone (due to only
the quadrupole moment of the binary) do not account for the obliquity intermittency in the Styx-t1
simulation and that frequency of the binary perturbations or proximity to the mean motion resonance 
are important.  

We note that the orbital period to binary period ratio for Styx is near 3. 
Two degree oscillations in the orbital inclination 
are apparent in the third from top panel in Figure \ref{fig:styx-nd} that we attribute to proximity
to the 3:1 orbital resonance, with
 $  3 n_o \sim n_B$.   Expansion of the disturbing function
gives a series of arguments that depend on $3 \lambda_o - \lambda_B$ and also include some
angles $\varpi_B, \varpi_o, \Omega_B, \Omega_o$  (e.g., \citealt{M+D}).  
Here $\lambda_B, \varpi_B, \Omega_B$
are the mean longitude, longitude of pericenter and longitude of the ascending node of the binary
and $\lambda_o,\varpi_o, \Omega_o$ are similar orbital elements but for the satellite.
The binary induced $J_2$ causes $\dot \Omega_o$ to be negative and $\dot \varpi_o$
to be positive (see equations \ref{eqn:dot_Om}, \ref{eqn:dot_varpi}).
The resonance term with argument 
\begin{equation}
\phi_{o2} = 3 \lambda_o - \lambda_B - 2 \Omega_o  \label{eqn:phi_o2}
\end{equation}
is sometimes called the $I^2$ resonance as it is second order in orbital inclination 
(e.g., section 8.12 of \citealt{M+D}).  
It has frequency $\dot \phi = 3 n_o - n_B - 2 \dot \Omega_o$ giving commensurability where
\begin{equation}
\frac{P_o}{P_B} \sim 3 - 2 \frac{\dot\Omega_o}{n_o}.
\end{equation}
As $\dot \Omega_o <0$ this implies that the inclination sensitive resonant subterm is encountered
at $P_o/P_B > 3$, whereas as $\dot \varpi_o >0$ the eccentricity subterms are encountered at $P_o /P_B <3$.
Styx, with $P_o/P_B \approx 3.15$, is near the inclination sensitive part of the resonance 
as $\dot \Omega_o/n_o \sim -0.015$ (using equation \ref{eqn:dot_Om} and Styx's value
for $\mu_B/M_B (a_B/a_o)^2$ from Table \ref{tab:sats}).     The oscillations in orbital
inclination in Figure \ref{fig:styx-nd} are from that resonance.

Are the intermittent  obliquity variations caused by proximity to the 3:1 resonance, and in particular
proximity to the inclination sensitive part of the resonance and so on the sign of $P_o/P_B -3$?
We ran an additional simulation Styx-t4, similar to the Styx-t1 simulation but with a somewhat wider binary.
This simulation has larger eccentricity oscillations because the system is outside the 3:1 resonance, 
rather than inside it. $P_o/P_B -3$ is negative rather than positive, 
and so the satellite is near the eccentricity sensitive region of the resonance rather than
the inclination sensitive region.
  The trajectory of the spin/mean-motion ratio exhibits 
some waviness but there are no strong obliquity 
variations.    We also ran a simulation identical to the Styx-t1 simulation but starting
with zero orbital inclination.
This simulation exhibits few degree variations in inclination, consistent with 
proximity to the inclination sensitive part of the 3:1 mean motion 
 resonance but lacks intermittency and large swings in the obliquity.   
The obliquity intermittency
seen in the Styx-t1 simulation requires proximity to the inclination sensitive parts of the
 3:1 orbital resonance with the binary and a few degrees of orbital inclination.
 
\subsection{Kerberos}

In terms of the ratio $w/n_o$, Kerberos is spinning about as fast as Styx.  We
show a simulation for its spin-down evolution 
 in Figure \ref{fig:ker_nd}.   Two simulations are
shown, the first (Ker-t1) with a slower rate of tidal dissipation than the second.  
The Ker-t1 simulation (Figure \ref{fig:ker_nd}a)
shows a short spin resonance capture in the middle the simulation, evident from the leveling
of $w/n_o$ in the second-left panel at $t \sim 1.8\times 10^6$.  
During the event, the body started tumbling
(see bottom right two panels) and the body was not stable in attitude; the obliquity increased
by about $5^\circ$.   
Inspection of third right panel shows that the spin itself was not exactly constant, in fact the
spin increased while the obliquity increased.   
Also plotted in the third-right panel (in orange) is the component of the spin vector in the direction
of the maximum principal axis.  While the spin increases, this component on average is level, 
implying that  a resonance maintained the magnitude of this component rather than the total spin.
The tumbling episode  appears to be associated
with a spin-orbit resonance as this occurred at $w/n_o \sim 6$. 
However, as we will discuss below
this resonance should be extremely weak due to the low orbital eccentricity.
The third from top and right panel shows that we failed to find a spin-binary resonance
in the form given by equation \ref{eqn:kB} at this spin value.
Below (section \ref{sec:wobble}) 
we consider the possibility that the resonance involved the precession rate of a wobbling
or tumbling body.    

The spin down rate was higher in the Ker-t2 simulation (Figure \ref{fig:ker_nd}b)
and the spin dropped to $w/n_o \approx 4$ where it appeared to be captured into spin-orbit resonance.
This spin value is much lower than Kerberos's current spin rate, but we include
the simulation here to illustrate a long lived spin-resonance capture event where
the obliquity increased  significantly, in this case to $60^\circ$. 
 The low orbital eccentricity again implies that
the spin orbit resonance should be very weak.  
The ratio $w/n_o$ is slightly lower than 4 suggesting that the resonance
is not a spin-orbit resonance but another type, though again there is no spin-binary resonance 
in the form given by equation \ref{eqn:kB} at this spin value.
A similar simulation that was run longer than the Ker-t2 simulation
showed that the body eventually
escapes the resonance, leaving the obliquity near $85^\circ$.

Spin-binary resonances were crossed in both Ker-t1 and Ker-t2 simulations with indices
$k_B = 2,3$ (see equation \ref{eqn:kB}) giving kinks in the spin rate plots as they were crossed.
Kerberos's spin rate is similar to the binary mean motion, 
and these resonances are among the strongest of the spin-binary resonances
(see \citealt{correia15}).
These two simulations show that
long lived spin-orbit resonance or spin-binary resonance capture 
(and associated large obliquity variation) are
 unlikely for $w/n_o \sim 6$ at Kerberos's current spin value.
 
Because the spin to mean motion ratio $w/n_B$ for
Styx and Kerberos are similar, the Styx and Kerberos simulations listed in Table \ref{tab:series_nd}  
have almost identical parameters.
The body axis ratios are also similar.  The primary difference between the Styx-t1 and Ker-t1,t2 simulations
is in the binary semi-major axis (compared to the orbital semi-major axis) placing Styx near the 3:1
mean motion resonance and Kerberos near the 5:1 resonance. 
Relative to the orbital semi-major axis, 
the binary semi-major axis is larger for the Styx simulations than
the Kerberos simulations and so the binary quadrupole moment 
 is larger and the binary is a stronger perturber.   The 
secular precession frequencies (in the longitude of ascending node) induced by the binary quadrupole
also differ in the two simulations.  We ran a series of simulations for Kerberos varying the
binary semi-major axis ratio and proximity to and side of the 5:1 resonance, but none
exhibited the obliquity intermittent variations of the Styx-t1 simulation. 

\subsection{More on Nix}

The simulation Nix-t1 has orbit to binary period ratio slightly smaller than 4, similar to that observed.
In contrast the orbit to binary period ratio of Styx is greater than 3, and we suspect that obliquity
variation in the Styx-t1 simulation is related to the proximity of the 3:1 inclination resonance with the
binary.  We are curious to see if a Nix simulation on the other side of the 4:1 resonance might exhibit
larger obliquity variation.   The Nix-t2 simulation is similar to the Nix-t1 simulation but has a smaller
binary semi-major axis.  This simulation exhibits obliquity oscillations of about 6 degrees in amplitude
that seem to be coupled with variations in inclination.  However large swings in obliquity are not
seen and spin-orbit and spin-binary resonances have no visible effect on the satellite spin 
when they are crossed. 

\vskip 0.1 truein

\subsection{Spin-orbit resonances}

Inspection of the second from top and left most panels in 
Figures \ref{fig:styx-nd} - Figure \ref{fig:ker_nd} shows that spin-orbit resonances 
predominantly do not cause large jumps in spin.
At the high spin rates of Pluto's minor satellites the spinning body 
is not likely to be captured into one of them.  
Spin orbit resonance is most simply described 
for a satellite spinning about a principal body axis that is aligned with the orbit normal.
The orientation of the satellite's long body axis is specified with angle $\vartheta$
and $\vartheta - f$ specifies the orientation of the satellite's long axis with respect to 
the planet-satellite center line, with $f$ the orbital true anomaly.
Expanding in a Fourier series, the equation of motion  
\begin{equation}
\frac{d^2 \gamma}{d t^2} +  \frac{n_o^2 \alpha^2}{2} \sum_p H(p, e) \sin (2 \gamma) = T
\end{equation}
\citep{goldreich68,wisdom84},
where $\gamma = \vartheta - p \lambda$ with $\lambda$ the orbital mean longitude, $p$ is a half integer and $\alpha$ is the asphericity dependent on body shape (equation \ref{eqn:alpha}).
The time averaged torque from tides (averaged over the orbit) is $T$ and $\alpha$ is
the asphericity defined in equation \ref{eqn:aspher}.
The coefficient $H(p,e)$ is a power series in orbital eccentricity $e$ and for $p>1$, the coefficient 
$H(p,e) = O\left( e^{p-2} \right)$ 
(\citealt{cayley1859}; also see \citealt{celletti} Table 5.1).
The $p$-th spin orbit resonance width can be described with the frequency  
\begin{equation}
\omega_{so,p} = n_o \alpha \sqrt{  H(p, e)}. \label{eqn:lib_so}
\end{equation}
 \citep{wisdom84}.
This frequency also characterizes the size of a jump in spin for a system crossing the resonance
and influences the likelihood of resonance capture.
If $p$ is large and the eccentricity is low then the resonance is narrow and weak
and the system unlikely to be captured into resonance.
Pluto and Charon's minor satellites have $w/n_o$ spin to mean motion ratios greater than 6
requiring half integer index $p>12$ and resonance widths $\propto e^5$ or higher.  
We can attribute the unimportance of the spin-orbit resonances to the high satellite spins,
requiring high orders in eccentricity for the spin-orbit resonance strengths 
and making them  weak.

\subsection{Spin-binary Resonances}

\citet{correia15} identified spin-binary resonances where
\begin{equation}
w - n_o = \frac{k_B}{2} (n_B - n) \label{eqn:binres}
\end{equation}
with integer $k_B$.
We have slightly rearranged equation \ref{eqn:kB}  so that it is clearer
that in a frame rotating with the satellite in its orbit (at angular rotation rate $n_o$)
the satellite body rotation rate is a multiple $k_B/2$ of the binary rotation rate.
\citet{correia15} estimated the strengths of the lowest order (in eccentricity) spin-binary resonances
of this form.  General expressions for resonant potential interaction terms are given in terms of Hansen coefficients 
(see equation 6 by \citealt{correia15}).
The derivations by \citet{correia15} could  in future
be generalized to depend on additional angles (all three Andoyer-Deprit angles), 
however the
complexity of the gravitational potential \citep{ashenberg07,boue17} 
makes this a daunting prospect.

Taking only the $k_B = 2,3$ terms in their equation 7,  and with 
an unequal mass binary, 
the spin-binary resonance libration frequencies are 
\begin{eqnarray}
\omega_{sb,k_B=2} &\approx & \left[ \frac{3}{8} \frac{ \mu_B}{M_B} \left(\frac{a_B}{a_o}  \right)^2 \right]^\frac{1}{2} n_o \alpha   \label{eqn:kb2} \\
\omega_{sb,k_B=3} &\approx & \left[ \frac{5}{16} \frac{ \mu_B}{M_B} \left(\frac{a_B}{a_o}  \right)^3 \right]^\frac{1}{2} n_o \alpha  . \label{eqn:kb3}
\end{eqnarray}
 The libration frequencies (and resonance strengths) do not strongly depend on orbital
eccentricity, as do spin-orbit resonances, 
and their resonance strength
falls only weakly with orbital radius or semi-major axis.  As a consequence 
spin-binary resonances might be fairly strong in the Pluto-Charon minor satellite system.

The dependence on semi-major axis of the spin-binary libration frequencies 
 (and the commensurability equation \ref{eqn:binres}) 
 implies
that the $k_B=2$ spin-binary resonance arises from the quadrupole gravitational moment of
the binary.   The $k_B=3$ spin-binary resonance strength is probably 
from the binary's octupole gravitational moment.
Figure \ref{fig:ker_nd}b, showing the  Ker-t2 simulation, illustrates that the $k_B=3$ resonance
is somewhat weaker than the $k_B=2$ resonance, as we would expect from the higher power
dependence on the semi-major axis ratio for the $k_B=3$ resonance.
Based on dependence on semi-major axis for the $k_B=2,3$ spin-binary resonances,
we expect that the larger $k_B$ spin-binary resonances consecutively depend on higher order
gravitational moments.   Since each gravitational potential moment gains a factor of the semi-major
axis and the libration frequency depends on the square root of the perturbation potential energy, 
we guess that the libration frequency depends
on $k_B$ as 
\begin{eqnarray}
\omega_{sb,k_B} \approx \sqrt{ \frac{ \mu_B}{M_B}}  \left( \frac{a_B}{a_o}  \right)^\frac{k_B}{2} n_o \alpha .
\end{eqnarray}
The increased sensitivity to distance from the binary makes the higher $k_B$ spin-binary resonances
weaker but perhaps not as weak as high index spin-orbit resonances that depend on 
high powers of orbital eccentricity. 
At higher satellite spin values, the only nearby spin-binary resonances have
higher $k_B$ indices.  The Nix simulation Nix-t1 shown in Figure \ref{fig:nix-nd}
shows the body crossing a $k_B$ = 9, 10 and 11 resonances with no effect on
any quantity we measured during the simulation.  
The $k_B = $ 5, 6  spin-binary resonances might have affected Styx's spin  in the 
Styx-t1 simulation
with a small jumps in spin when they were crossed (see Figure \ref{fig:styx-nd}, third-right panel).
The weakness of spin-binary resonances at Nix's higher spin, compared
to their relative strength in Kerberos's simulations  is consistent with
our hypothesis that the higher $k_B$ spin-binary resonances are weaker than the
lower ones.    A consequence is that spin-binary resonance overlap is not assured
at the higher observed satellite spins. The binary mediated mechanism causing
tumbling proposed by \citet{correia15} may not operate at the higher spin values.

Pluto and Charon's minor satellites have asphericity 
$\alpha \sim 1$, mass ratio $\mu_B/M_B \sim 0.1$, and semi-major axis ratio
$\sim 2.5$  giving
$\omega_{sb,k_B=2}/n_o \sim 0.1 $, far exceeding the high $p$ spin-orbit resonance strengths
that depend on high powers of the eccentricity.
Kerberos currently has ratio of binary to spin period of 1.2 and so is near
the $k_B=2$ resonance and Figure \ref{fig:ker_nd}b shows evidence of
this resonance affecting satellite spin with a small jump in spin as this resonance was crossed.
The jump in spin was about $\delta w = 0.02$, at  $w \sim 0.33$ and $w/n_o \sim 5 $ giving 
$\delta w/n_o \sim 0.02 \times 5/0.33 \sim 0.3 $.   
The jump size in spin should be  and is approximately the same size as the estimated libration frequency,  and as expected.

\begin{table*}
\vbox to45mm{\vfil
\caption{\large  Parameters for simulations with a migrating binary\label{tab:series_b}}
\begin{tabular}{@{}llllllll}
\hline
Simulation name & & Styx-b1 & Styx-b2 & Nix-b1 & Nix-b2  \\
\hline
Spring damping rate   & $\gamma_s$ &0.001 & 0.001 & 0.1 & 0.1  \\
Tidal frequency          & $\bar \chi$   & $2 \times 10^{-4}$ & $2 \times 10^{-4}$ & 0.04 &0.04   \\
Initial obliquity & $\epsilon_{init}$  & $5^\circ$ & $20^\circ$ &  $5^\circ$ & $20^\circ$\\
Initial binary semi-major axis & $a_B$ & 275 &  275 & 324 & 324 \\
\hline
\end{tabular}
{\\  For these simulations the binary mass ratio $q_B = 0.12$  and the binary 
 semi-major axis drift rate  $\tau_a^{-1} = 5\times 10^{-8}$ . 
}
}
\end{table*}

\begin{figure*}
\includegraphics[width=5in]{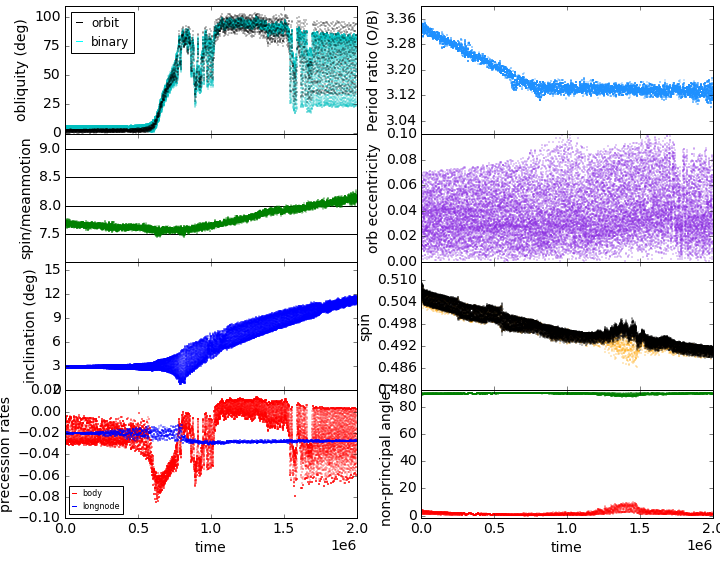}
\includegraphics[width=5in]{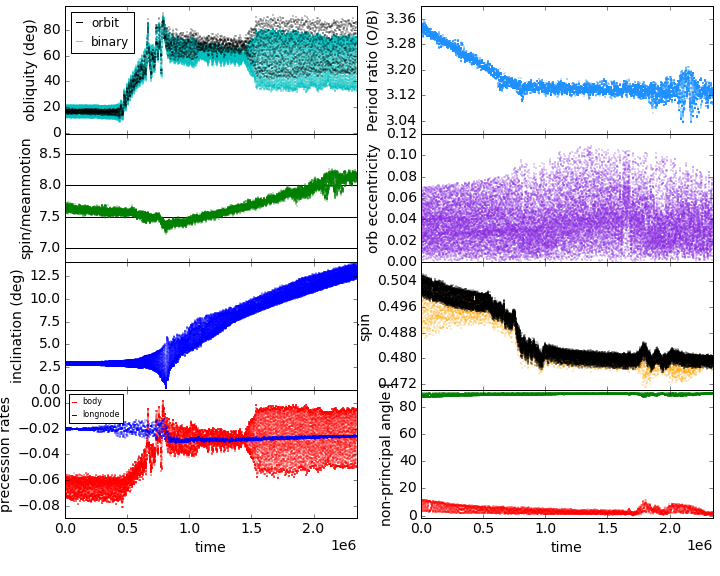}
\caption{Simulations of Styx with a slowly separating binary.  
a) The Styx-b1 simulation 
starting at an obliquity of $5^\circ$.  
b) The Styx-b2 simulation, 
starting at an obliquity of $20^\circ$.  
Simulation parameters are
listed in Tables \ref{tab:common}, \ref{tab:series_body} and \ref{tab:series_b}.
Slow separation of the binary captures the minor satellite into 3:1 mean motion resonance
which lifts the orbital inclination.   The obliquities are lifted to near $90^\circ$. 
\label{fig:styx-b}}
\end{figure*}

\begin{figure*}
\includegraphics[width=5in]{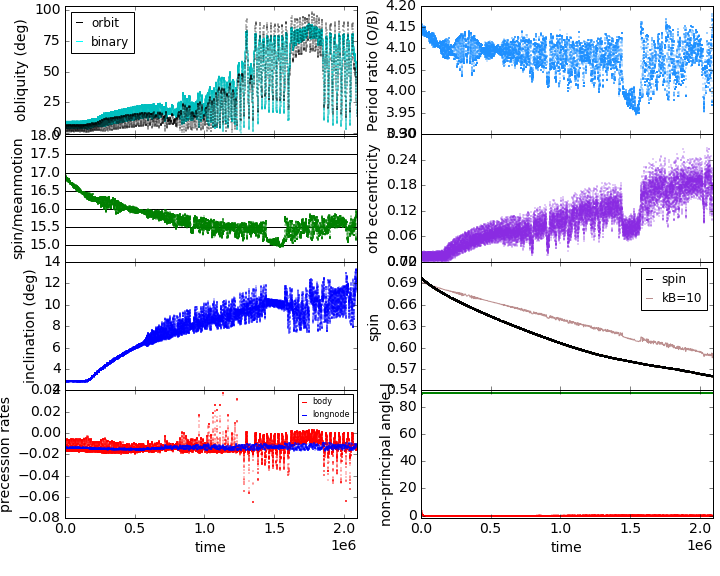}
\includegraphics[width=5in]{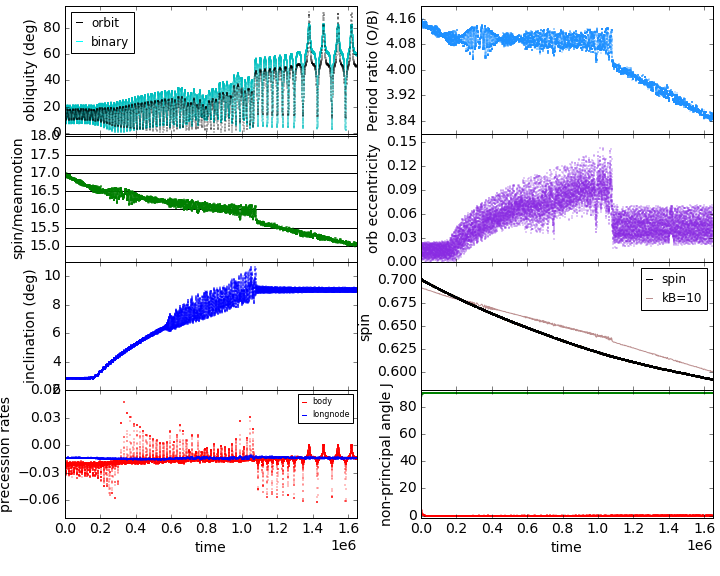}
\caption{Simulations of  Nix with a slowly separating binary. 
a) The Nix-b1 simulation 
starting at an obliquity of $5^\circ$.  
b) The Nix-b2 simulation, 
starting at an obliquity of $20^\circ$.  
Slow separation of the binary captures the minor satellite into 4:1 mean motion resonance
which lifts the orbital inclinations and obliquities.  
Simulation parameters are
listed in Tables \ref{tab:common}, \ref{tab:series_body} and \ref{tab:series_b}.
\label{fig:nix-b}}
\end{figure*}

\subsection{Excitation of Wobble} \label{sec:wobble}

As some resonant features have not yet been identified, 
we consider excitation of wobble, where the spin axis and body principal axis are not aligned.
In this setting,
a torque free axisymmetric body has body principal axis and spin axis both
precessing about the spin angular momentum vector.  With all axes nearly aligned
a precession or wobble frequency as seen by an inertial observer is
\begin{equation}
\frac{\dot \varphi}{w} \sim \frac{I_\parallel}{I_\perp}  
\end{equation}
where $I_\parallel$ is the moment of inertia about the body's symmetry axis 
and $I_\perp$ is that around an axis perpendicular to the symmetry axis.
Here $\varphi$ is the Euler precession angle measured in a coordinate system
aligned with $z$ axis in the direction of the angular momentum vector.

Pluto and Charon's satellites are not axisymmetric, however they are spinning rapidly.
We approximate the precession rate by replacing $I_\perp$ with an average value $(B+A)/2$
and setting $I_\parallel =C$ where $C\ge B\ge A$ are the body's moments of inertia;
\begin{equation}
\frac{\dot \varphi}{w} \sim \frac{2C}{A+B}  \label{eqn:ratio}.
\end{equation}
In analogy to equation \ref{eqn:binres} for the spin resonances associated with the binary
we guess that tumbling could be excited where
\begin{equation}
\dot \varphi - n_o = j_\varphi (n_B - n_o). \label{eqn:phires}
\end{equation}
We plotted the location for $w$ for a tumbling resonance with index $j_\varphi=2$ (using
equation \ref{eqn:phires} and \ref{eqn:ratio})  as a green horizontal line on 
Figure \ref{fig:ker_nd}a.  
We found that the tumbling episode at $t=1.8 \times 10^6$ in the Ker-t1
simulation is likely caused by excitation of tumbling with a binary perturbation 
frequency matching  the wobble or precession frequency.  We searched for 
but failed to find a similar
resonance accounting for the resonance seen at $t>0.5 \times 10^6$ in Figure \ref{fig:ker_nd}b.

\section{Simulations with an outwards migrating binary}  \label{sec:bin}

With tidal dissipation alone and a moderate level of spin-down we can only explain
the high obliquity of Styx.
We now explore the possibility that the satellites could have been captured
into mean-motion resonance due to outward migration of Charon with respect to Pluto.
Slow separation of the Pluto-Charon binary could have
 been caused by tidal interaction between Pluto and Charon \citep{farinella79}.  Mean motion
 resonant capture can also take place if a minor satellite
 migrates inward, and this could have occurred via
 interaction with a circumstellar disk.

To migrate the binary (slowly separate Pluto and Charon) 
we apply small velocity kicks to each body in the binary using the recipe for
migration given in equations 8-11 by \citet{beauge06}. The kicks are applied so as to keep
the center of mass velocity fixed.  The migration rate, $\dot a_B$, 
depends on an exponential timescale
 $\tau_a$ (the parameter $A$ in the equation 9 by \citealt{beauge06}).  The migration
 rate depends on $\tau_a^{-1}$ with $\dot a_B \sim a_B \tau_a^{-1}$.
 We adopt a convention $\tau_a >0$ corresponding to outward migration which allows
an external minor satellite to be captured into mean-motion resonance.

We ran
a series of simulations with a migrating binary and with parameters listed in Tables \ref{tab:common},  \ref{tab:series_body},  and \ref{tab:series_b} that are shown in 
Figures \ref{fig:styx-b} -- \ref{fig:nix-b}.
Initial conditions
 are similar to those listed in Table \ref{tab:series_nd} except
the initial binary semi-major axis is smaller so as to let the
binary approach the current satellite values.  The body axis ratios are
identical, but the dissipation in the spinning body is lower (reducing tidal drift), 
and the initial obliquities are $5^\circ$.

Figure \ref{fig:styx-b} -- \ref{fig:nix-b} show that as the binary separates,
the minor satellite captured into mean motion resonance. 
In the Styx-b1,b2 simulations
it is the 3:1 mean motion resonance and initially only the inclination increases, 
whereas for the Nix simulations the mean motion resonance is the 4:1 and both eccentricity
and inclination increase. The 4:1 resonance is a third order resonance (in eccentricity)
and its lowest order resonant arguments all contain the longitude of pericenter, $\varpi$ 
(see the appendices by
\citealt{M+D}).
All resonant subterms affect the eccentricity.
The 3:1 resonance
is second order and does contain subterms with arguments that lack the longitude of pericenter
and so one of these causes an increase in the inclination and not the eccentricity.
 In both cases the minor satellite obliquity is lifted to high values,
near $90^\circ$.   The lift in obliquity for Nix is particularly interesting
because it is a mechanism for lifting obliquity that functions
even at Nix's high spin rate.  The mechanism also works for Styx even
though we found previously that Styx can undergo intermittent obliquity 
without mean motion resonance capture. 

We ran similar simulations for Kerberos drifting the binary apart,
but none of our simulations illustrated capture into 5:1 mean motion resonance.
Kerberos is more easily captured into mean motion resonance by Nix and that could
have lifted its orbital inclination.  We suspect that a similar obliquity lifting 
mechanism might work for Kerberos but it would involve at least four  bodies; 
Nix to capture Kerberos into mean motion resonance and a simultaneous commensurability
with Kerberos's spin precession rate and the Pluto-Charon binary to lift Kerberos's obliquity.

\begin{figure*}
\includegraphics[width=3in]{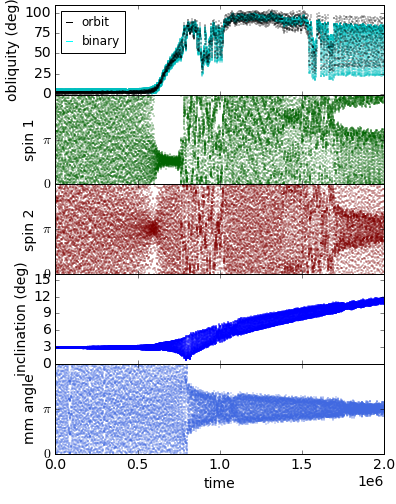}
\includegraphics[width=3in]{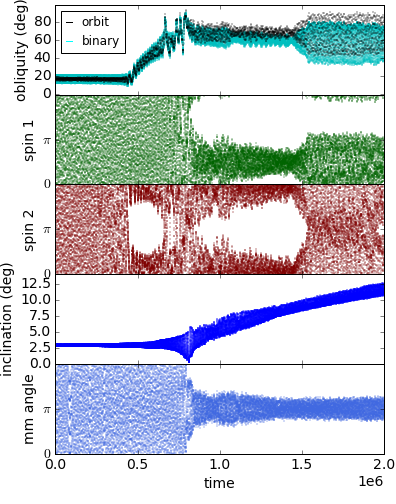}
\caption{Resonant arguments for simulations of Styx with a slowly separating binary.  
a) The Styx-b1 simulation (also shown in Figure \ref{fig:styx-b}a) with quantities plotted as a function of time. 
The top panel shows obliquity in degrees.  The second panel shows 
the argument $\phi_{s1} = 3 \lambda_o - \lambda_B - \Omega_s -\Omega_o$ in radians.
The third panel shows the argument $\phi_{s2}= 3 \lambda_o - \lambda_B - 2\Omega_s$ in radians.
The fourth panel shows orbital inclination in degrees.
The bottom panel shows argument $\phi_{o2}=3 \lambda_o - \lambda_B - 2\Omega_o$ in radians.
b) The Styx-b2 simulation (also shown in Figure \ref{fig:styx-b}b. 
Angles involving the precession angle $\Omega_s$ of the spinning satellite
are approximately fixed when the obliquities increase.  This occurs prior
to entering the mean motion resonance.  The satellite is in the mean-motion resonance
 when the inclination increases and $\phi_{o2}$ librates about $\pi$.
\label{fig:styx-ang}}
\end{figure*}

\subsection{Obliquity increase near mean-motion resonance}

Because Nix is spinning more rapidly than Styx, its spin precession rate, $\dot \Omega_s$,  
is closer to $\dot \Omega_o$,
its precession rate of the longitude of the ascending node.
The Nix-t1 simulation (Figure \ref{fig:nix-nd}a)
 begins with the spinning body in a Cassini state with $\dot \Omega_o \approx \dot \Omega_s$
and the system remains in a Cassini state except for a brief period near $t=1.7 \times 10^6$.
That $\dot \Omega_o \approx \dot \Omega_s$ means that in mean motion resonance the spin
precession rate
lines up with the mean motion resonant angle and this coupled with the binary inclination is likely to
account for the large obliquity variations.  
In resonance the binary perturbations are in phase with the tilt angle of the body.
A similar simulation but with an initial obliquity of $20^\circ$, the Nix-t2 simulation,  (Figure \ref{fig:nix-b}b) 
starts outside Cassini
state ($\dot \Omega_s \ne \dot \Omega_o$) 
but moves into it after mean-motion resonance capture, probably because of the orbital 
inclination increase caused by the mean motion resonance.  
 In that simulation the satellite exits mean motion resonance leaving
the body in a state similar to that at the end of the Nix-b1 simulation
(Figure \ref{fig:nix-b}a) with high obliquity oscillations.

The Styx-b1 simulation (Figure \ref{fig:styx-b}a) also begins in a Cassini state.  
As the system approaches mean motion resonance
there is a large increase in obliquity.   Mean motion resonance is entered about $t=0.8 \times 10^6$ whereas
the obliquity increase begins at about $t=  0.6 \times 10^6$ at which time
the satellite also exits Cassini state.   Outside of the Cassini state the body spin precession rate is 
about 3 times higher
(in amplitude) than the rate of precession of the longitude of the ascending node.  
As the mean motion resonance is approached the body spin precession rate 
would be commensurate with
the rate of change of the mean motion resonance angle 
before the longitude of ascending node is commensurate.  In other words  
$2 \dot \Omega_s \sim 3 n_o - n_B$ prior to $2\dot \Omega_o \sim 3 n_o - n_B$.
The increase in obliquity prior to entering the mean motion resonance is more clearly
seen in the Styx-b2 simulation (Figure \ref{fig:styx-b}b) as it only enters Cassini state after capture
into the mean motion resonance.

We suspect that the obliquity increases prior to entering mean motion resonance 
are due to a subresonance involving the Euler angle $\Omega_s$.  
To check this possibility we plot inclination and obliquity for the Styx simulations along with 
resonant angles in Figure \ref{fig:styx-ang} for simulations Styx-b1 and Styx-b2.
In these figures three resonant arguments are plotted:
\begin{eqnarray}
\phi_{s1}  &=& 3 \lambda_o - \lambda_B -  \Omega_s  - \Omega_o \nonumber \\
\phi_{s2}  &=&  3 \lambda_o - \lambda_B - 2 \Omega_s  \nonumber \\
\phi_{o2} &=& 3 \lambda_o - \lambda_B - 2 \Omega_o, \label{eqn:arguments}
\end{eqnarray}
where $\lambda_o, \lambda_B$ are the mean longitude of satellite and binary, respectively,
$\Omega_o$ is the longitude of the ascending node of the satellite and
$\Omega_s$ is the precession angle of the spinning satellite.
The last of these arguments, $\phi_{o2}$,  is the argument associated 
with the $I^2$ (inclination squared) part of the 3:1 mean motion resonance.
The top two arguments involve the precession angle of the satellite.  Figure \ref{fig:styx-ang}a, showing
the Styx-b1 simulation,  shows $\phi_{s1}$ freezing (or librating about a fixed value) 
 during the same time period that the obliquity
increases, whereas  Figure \ref{fig:styx-ang}b  showing
the Styx-b2 simulation,  shows $\phi_{s2}$ freezing (librating about 0) 
during the time period that the obliquity
increases.  The freezing of these angles in the simulations suggests that these resonances are responsible
for the large obliquity variations.
In both Styx-b1,b2 simulations, $\phi_{o2}$, associated with the mean motion resonances,
 is only librating when the orbital inclination increases and after the obliquity has reached
a high value.

The simulations shown in Figures \ref{fig:styx-b} - \ref{fig:styx-ang} with an outward drifting
binary suggest that obliquity variation is associated with an increase in orbital inclination and
proximity or capture into mean motion resonance.  
However the obliquity increases tend to take place
just before entering resonance
implying that a subresonance involving the spin precession is responsible.
Since this type of resonance is associated with the spin precession we could call it a three-body 
secular resonance,
except with such elongated bodies as Pluto's minor satellites the spin precession frequency 
at low obliquity is not particular slow.   And since such a commensurability involves a mean motion resonance,
in terms of its orbital properties it is not secular (it depends on mean longitudes which are usually
averaged for secular resonances).  
The resonance (involving
spin precession and mean motions) might be important precisely because
 the spin precession frequencies are fast.
 

The argument $\phi_{s2}$ in equation \ref{eqn:arguments} can be written
\begin{eqnarray} \phi_{s2} &=& 3 \lambda_o - \lambda_B - 2 \Omega_s \nonumber \\
&=& ( \lambda_o - \lambda_B) + 2( \lambda_o - \Omega_s) . \label{eqn:phis2_o}
\end{eqnarray}
In a frame moving with the spinning body (here Styx) in its orbit, but not rotating with the spinning body,
angles are subtracted by $\lambda_o$.  This picture is similar to those used to explore Lindblad resonances.
In this frame, and with $\phi_{s2}$ nearly constant, the Pluto-Charon binary appears to rotate
twice during the same time that Styx precesses a full periods. 

We lack a model for resonances with arguments given by $\phi_{s1}, \phi_{s2}$ in equation 
\ref{eqn:arguments}, but we can estimate a  timescale associated with obliquity increase in resonance.
We suspect that a torque in resonance would be a few times lower than 
the torque in a spin-binary resonance --  
 a few times lower because we must average the effect
of the spin-binary resonance over the orbit while in mean motion
resonance and because the strength also  depends on the obliquity.  
Ignoring these dependencies (and using equation \ref{eqn:kb2}) the torque
in our spin resonance is of order  
$T \sim n_o^2 \alpha^2 \frac{\mu_B}{M_B} \left(\frac{a_B}{a_o} \right)^2 I$
with moment of inertia $I$.  We estimate a timescale for obliquity change with $ t_{obl} \sim Iw/T$ giving
\begin{equation}
t_{obl} n_o \sim  \frac{w}{n_o} \left[ \left(\frac{\mu_B}{M_B}\right)  \left(\frac{a_B}{a_o} \right)^2 \right]^{-1}. \label{eqn:tobl}
\end{equation}
Taking $w/n_o \sim 6$ and $\frac{ \mu_B}{M_B} \left(\frac{a_B}{a_o} \right)^2 \sim 0.01$ 
(from the bottom
of Table \ref{tab:sats}) we estimate a timescale
for a large obliquity change
$t_{obl} n_o \sim 600$ or about 100 orbital periods.  In units of $t_g$ (like our simulations)
an orbital period is about 100 giving $t_{obl} \sim 10^4$. 
The timescale for obliquity change seen for Styx (see Figures \ref{fig:styx-b}) is
about $10^5$ and for Nix, with higher spin, is slower a few times $10^5$ (see Figures \ref{fig:nix-b}).
This an order of magnitude higher than estimated with equation \ref{eqn:tobl}.
The discrepancy is comfortably wide, wide enough that drift within resonance, reduction in strength
from averaging over fast angles and body angular orientation (obliquity) and inclination 
dependence can probably be taken into account still giving
the resonance enough strength to lift the obliquity.  
As long as the drift rate is adiabatic, the system
can be captured into and stay in resonance.  In analogy to mean motion resonance capture, 
a canonical momentum variable, dependent on obliquity, 
rises to maintain the resonant angle, with
 rise time set by the drift rate rather than the resonance strength.
The condition for adiabatic drift is dependent on the resonance libration frequency \citep{quillen06}
with slower drifts than the adiabatic limit allowing resonant capture and evolution.
Our rough comparison between estimated and numerical obliquity rise times
suggests that this type of resonance is capable of lifting the obliquities. 

Our simulations do not exhibit obliquity $\epsilon_B$, with respect to the binary orbit,
greater than $90^\circ$, corresponding to retrograde spin.  
Styx and Kerberos have near $90^\circ$ obliquities whereas Nix and Hydra have higher
obliquities of 123$^\circ$ and 110$^\circ$, respectively.  
We notice from Table \ref{tab:sats}
that Styx and Kerberos have period ratio $P_o/P_C - j >0$ where $j$ is the nearest
integer (3 and 5 respectively), 
whereas Nix and Hydra have period ratio subtracted by the nearest integer $j$ (4 and 6 respectively)
less than zero.  
Styx and Kerberos have
 $j \dot  n_o - \dot n_B <0$ whereas $j \dot n_o - \dot n_B >0$ for Nix and Hydra.
 The two satellites with the highest, and retrograde obliquities also have positive $j \dot n_o - \dot n_B$.
Perhaps  there is a connection between the obliquity and the side of the orbital resonance.
With retrograde spin (or $\epsilon_B > 90^\circ$), the spin precession rate $\dot \Omega_s >0$ rather
than 
negative (as is true for Styx and Kerberos).  
Thus Nix and Hydra could be near commensurability with fixed or librating resonant argument $\phi_{2s}$
or $\phi_{1s}$. 

With initial conditions at low obliquity, our simulations
did not ever exhibit retrograde obliquities, but perhaps with additional orbital migration
retrograde spins could be induced.  The one simulation where the system leaves mean motion resonance
and crosses to the side of mean motion resonance that Nix and Hydra are on
is the Nix-b2 simulation (Figure \ref{fig:nix-b}b). However, in this simulation, 
the satellite did not stay in spin resonance 
when exiting the mean motion resonance though the satellite did stay in a Cassini state.
The spin state exhibits high obliquity swings and 
perhaps further evolution could induce retrograde spin, 
or there may be a diversity of ways that the body can exit the mean motion
resonance in the full N-body system including all  minor satellites.
The exit state seen in Figure \ref{fig:nix-b}b is not unique.  Similar simulations
exited the mean motion resonance also at high mean obliquity but with different
amplitudes of oscillation about a mean value. 

\begin{figure*}
\includegraphics[width=3in]{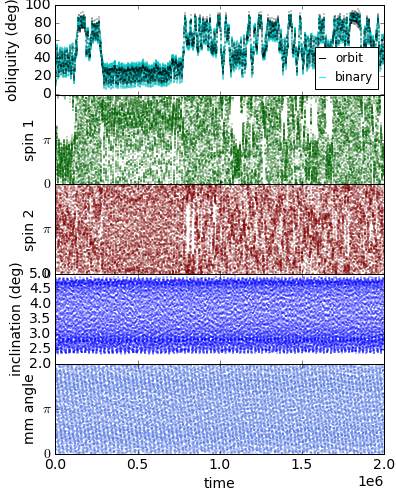}
\caption{Resonant arguments for the Styx-t1 simulation with tidal dissipation but lacking
binary migration.  Similar to Figure \ref{fig:styx-ang} but for the simulation shown in Figure
\ref{fig:styx-nd}.
Intermittent obliquity evolution may be caused by resonances involving arguments 
$\phi_{s1} = 3 \lambda_o - \lambda_B - \Omega_s -\Omega_o$ and
$\phi_{s2}= 3 \lambda_o - \lambda_B - 2\Omega_s$, shown in the second and third panels.
\label{fig:styx-nda-ang}}
\end{figure*}

 We reexamine the intermittent obliquity variations shown
in simulation Styx-t1 (Figure \ref{fig:styx-nd}) lacking binary migration.  
The obliquity panel (top left) shows that
when the obliquity is high, the body is in a Cassini state
and when the obliquity is low it is not in one.   When the obliquity is low, the precession rate is high enough
that it might be commensurate with the 3:1 mean motion resonant angle. 
To see if this is the case
we created a figure, Figure \ref{fig:styx-nda-ang}, similar to  Figure \ref{fig:styx-ang} 
but for the Styx-t1 simulation that lacks binary migration.
Resonant angles $\phi_{s1}, \phi_{s2}$ show librating regions and these do occur during
obliquity variations.
We might attribute the intermittent chaotic obliquity evolution to perturbations from multiple terms involving
the mean motion resonance and the body's spin precession angle.

The resonant mechanism seen here for obliquity increase differs from
the mechanism proposed by \citet{brunini06a} that was  retracted \citep{brunini06b}.
The scenario is as follows:
 When Jupiter and Saturn crossed the 2:1 mean-motion resonance, their orbital eccentricities were excited and the orbits of Uranus and Neptune were destabilized. This  caused a series  
of close encounters between the giant planets that generated high obliquities and giving
a possible scenario accounting for Uranus' high obliquity.
\citet{lee07} presented simple analytic argument showing why close encounters between
planets could not cause permanent and large obliquity variations.
The change in the spin direction of a planet relative to an inertial frame during an encounter 
between the planets is very small. The change in the obliquity 
is due to the change in the orbital inclination. 
As inclinations would be damped by planetesimal interactions on timescales shorter 
than the timescales on which the spins precess, the obliquities should return to small values if they are small prior to the encounters.

Our mechanism is based on a spin-resonance and does not require
close encounters.   The spin-secular resonance \citep{ward04}
is capable of causing large obliquity variations  and also does not
involve encounters.  The obliquity variations are not tightly coupled to orbital
inclination because a third body is involved.
Once captured into the spin resonance, adiabatic drift within resonance pushes
the spin to high obliquity and this is also the case for the spin-secular resonance.  
If the system exits the resonance slowly and in the same way
as it entered the resonance, the body would return to low obliquity.
However our simulations illustrate that the body can exit resonance abruptly rather
than adiabatically, leaving the spinning body at much higher obliquity than when it
entered the resonance (see Figure \ref{fig:nix-b}b).  In our setting the obliquity
and orbital inclination are not tightly coupled and this is because the inclination of
the binary varies, though very slightly as the binary is massive compared to the spinning
satellite.  Unless the system remains in the spin resonance, subsequent damping
of the orbital inclination by a circumbinary disk would not be expected to simultaneously
return the obliquity to its original value.    
The spin-resonant mechanism seen here, involving mean-motion resonance 
and an Euler angle that
can cause large obliquity changes, differs from the mechanism 
proposed by Brunini and should not suffer from the flaw \citep{lee07}
that invalidates Brunini's mechanism. 

\section{Summary and Discussion} \label{sec:sum}

Estimates of tidal spin-down time suggest
that none of Pluto and Charon's minor satellites have had time during the age
of the Solar system to reach near spin-synchronous states and this is consistent
with the high observed spin rates \citep{weaver16}.
As tidal evolution is slow, the minor satellites could have retained their primordial
obliquities.
If the minor satellites accreted from a circumbinary disk comprised of
small particles  then their primordial obliquities
would have been low.  Alternatively if they accreted from or suffered impacts with
massive bodies then their primordial obliquities would be randomly oriented.
The current obliquities of the minor satellites are all near enough to $90^\circ$ 
that the spins are unlikely to be consistent with
 four randomly oriented spins and so motivate a study of obliquity evolution
after formation.

Mass spring model
simulations with tidal dissipation, but allowing only moderate spin-down, 
show that only minor changes in minor satellite obliquity are caused
by crossing spin-orbit  resonances, though our simulation method
does exhibit spin-orbit resonance capture at lower spin rates. 
The high spin rates makes the spin-orbit resonances 
sensitive to orbital eccentricity to a high power and so very weak. 
Spin-binary resonances, as they depend on gravitational multipole moments
of the binary, are stronger  at the high spin rates.   
Nevertheless, only small jumps in spin are seen when crossing them.
 We have found that capture into them is rare, with the possible
exception of Kerberos that has spin rate similar to the binary mean motion. 
Only Styx experiences large and intermittent obliquity variations when evolving tidally.  
Proximity to the 3:1 mean motion resonance and few degree orbital inclination seem to be required
for Styx to show large and intermittent obliquity variations. 

Simulations allowing the Pluto-Charon binary to slowly drift apart cause Styx to be captured into
3:1 mean motion resonance with the binary and Nix to be captured into 4:1 resonance.
Inclination sensitive parts of these resonances are encountered first, and these increase the orbital
inclination.  The satellite obliquities are lifted to near $90^\circ$
either on approach to or in mean motion resonance, depending upon whether the satellite is
in a Cassini state or not.  We suspect that the obliquity increases are caused by
  a commensurability between the mean motion resonance argument  frequency and the 
satellite spin precession rate.  This resonance is likely because the satellites are
sufficiently elongated that the spin precession rates at low obliquity are fairly fast.
Re-examination of the Styx simulation showing intermittent obliquity variations suggests
that this type of resonance could contribute to Styx's chaotic behavior.
The mechanism for lifting obliquity, involving mean motion resonance and spin precession, functions
for both Nix and Styx even though Nix is spinning much faster than Styx.


We have explored only three-body integrations, the binary and a single resolved
elongated spinning body.  Our mechanism lifting 
Styx and Nix's obliquities was not effective in our simulations for Kerberos that failed
to capture into 5:1 mean motion resonance.  
Simulations involving 4 or more satellites might succeed in lifting Kerberos's obliquity with 
a similar mechanism.  Kerberos's orbital inclination could be increased via capture into 5:4 resonance with Nix 
and its obliquity  lifted at about the same time.
Since this mechanism is not strongly dependent on the satellite spin rate, it may also work
for the more rapidly spinning Hydra, perhaps via a 3:2 mean motion resonance with Nix.
As all four satellites are near mean motion resonances, a mechanism that involves
mean motion resonances and the body Euler precession 
angle and lifts obliquities (via matching spin precession)
 could operate effectively on all four minor satellites
perhaps explaining why all of them are near $90^\circ$.

Styx and Kerberos, both inside mean motion resonance with Charon,
have lower obliquities 
than Nix and Hydra, that are outside of mean motion resonance with Charon. 
There may be a connection between the direction of spin, prograde or retrograde, 
 and the side of mean motion resonance.
While our simulations did not induce retrograde spins, perhaps later orbital migration
or tidal evolution in the full
N-body system (with all 4 minor satellites) could induce these spin end-states.

We started our simulations with minor satellite spinning along a principal body axis, however
Kerberos's tumbling decay timescale might be long enough that it could experience orbital
evolution before its wobble decays.  In future we could explore spin evolution of initially
tumbling states; perhaps Kerberos is more likely to capture into spin resonances if
it is tumbling.

The current obliquities of the minor
satellites need not be near their primordial values if a spin resonance lifted them to high values.  
If the current high obliquities are related to mean motion resonances
then we could infer that all of the minor satellites were previous captured
into or in mean motion resonance.     However this would conflict with dynamical
studies of the orbital evolution showing that this causes instability \citep{cheung14}.
Perhaps migration, resonance capture and
 associated obliquity increases could have taken place when a circumbinary
disk was still present that could damp inclinations and eccentricities and stabilize the orbits --
or the system may have actually experienced episodes of instability and reformation.
Hybrid scenarios are also possible, for example with Styx experiencing
obliquity variations due to spin-binary or spin-mean motion resonance,
Kerberos experiencing obliquity variations due to spin-binary resonance
 and Nix and Hydra, at higher spin rates, retaining their primordial obliquities.
Investigation of these possibilities could explore whether high obliquities are
maintained as orbital inclinations are damped after exiting mean motion
resonance.

We lack simple dynamical models for phenomena seen in our simulations, such
as excitation of tumbling.
The complexity of the potential  (dependence on at least 3 angles) 
for quadrupole/quadrupole body gravitational interactions \citep{ashenberg07,boue17} implies that
constructing a more general model for spin-binary resonance
(beyond \citealt{correia15})
 that includes obliquity and tumbling  would be challenging.
Though we suspect a spin-precession/mean motion resonance mechanism 
for obliquity increase, we lack a dynamical model that would allow us to assess
the resonance strength.  

Our obliquity lifting mechanism seems to require past orbital misalignments with non-zero but few
degree satellite orbital inclinations with respect to the binary.   
Future observations or ongoing analysis of New Horizons observations
should determine whether such inclinations are currently ruled out.
After escape from resonance in the Nix-b2 simulation (Figure \ref{fig:nix-b}b) Nix 
exhibits large obliquity swings from near 0 to near $90^\circ$ at
a period of about $t=10^5$ or corresponding to order $10^3$ orbital periods or a few years
in real time.
If Nix or Styx is currently in such a spin state, large obliquity variations might be observed.

The obliquity lifting mechanism seen here involves fast precession rates 
(due to body elongation)
and mean motion resonance with a massive binary and does not
involve obliquity variations due to close encounters between bodies (as discussed by 
\citealt{brunini06a,brunini06b,lee07}).
 It would be interesting to explore other
settings where a similar spin resonance might operate.    Uranus is more nearly spherical
but in the past might have been in mean motion resonance with Saturn or Jupiter.
Perhaps Uranus's high obliquity could have been lifted because of its past interaction with
mean motion resonance, as seen here, rather than due to secular resonance \citep{rogo16}
or collisions (e.g., \citealt{parisi97}).  

Lastly higher quality and more accurate simulations can be carried out to 
check, confirm, correct and extend our findings.   
Simulations can be improved
by including all the minor satellites,
taking into account the shapes of Pluto and Charon and tidal dissipation in the binary, 
increasing the time integrated so that slower tidal evolution can be studied,
better resolving the spinning bodies with more particles and with correct
satellite radii,
using a  different type of rotational body simulation technique, 
taking into account perturbations due to the Sun and other planets,
using updated viscoelastic material properties and more complex simulated
rheology and 
using improved masses, shapes, orbital parameters and spins 
as observational measurements continue to improve upon measurements.

\vskip 1.5truein
Acknowledgements.

We thank David Trilling, Rob French, Alexandre Correia, Darin Ragozzine, and Mark Showalter
for helpful discussions and correspondence.
This work was in part supported by the NASA grant NNX13AI27G to ACQ.
BN is funded by the contract Prodex CR90253 from the Belgian Science Policy (BELSPO)"

\vfill\eject

{}

\end{document}